\documentclass[useAMS,usenatbib,fleqn]{mn2e}

\usepackage{subfigure}
\usepackage{graphicx, amssymb}
\usepackage[fleqn]{amsmath}
\usepackage{epsfig}
\usepackage{color}
\usepackage{times}
\usepackage{amsfonts}

\newcommand{\beq}{\begin{equation}}
\newcommand{\eeq}{\end{equation}}
\newcommand{\bea}{\begin{eqnarray}}
\newcommand{\eea}{\end{eqnarray}}

\title[Abundance anomalies in metal-poor stars]{Abundance anomalies in metal-poor stars from Population III supernova
ejecta hydrodynamics}

\author[A. Sluder et al.]
{Alan~Sluder$^1$, Jeremy S.~Ritter$^1$, 
Chalence~Safranek-Shrader$^{1,2}$, 
Milo\v  s~Milosavljevi\'c$^1$, \newauthor and Volker~Bromm$^1$\\
$^1$Department of Astronomy, University of Texas at Austin, Austin, TX 78712, USA \\
$^2$Department of Astronomy and Astrophysics, University of California, Santa Cruz, 95064, USA
 }
 
\newcommand{\apj}{ApJ}
\newcommand{\aj}{AJ}
\newcommand{\mnras}{MNRAS}
\newcommand{\araa}{ARA\&A}
\newcommand{\apjs}{ApJS}
\newcommand{\nat}{Nature}
\newcommand{\apjl}{ApJ}
\newcommand{\aap}{A\&A}

\renewcommand{\u}[1]{\,\textrm{#1}}
\newcommand{\f}[2]{\frac{#1}{#2}}
\renewcommand{\(}[1]{\left(#1 \right)}
\renewcommand{\[}[1]{\left[ #1\right]}
\renewcommand{\=}{&=&}

\begin{document}

\label{firstpage}

\maketitle

\topmargin-1cm


\begin{abstract}

We present a simulation of the long-term evolution of a Population III supernova remnant in a cosmological minihalo. Employing passive Lagrangian tracer particles, we investigate how chemical stratification and anisotropy in the explosion can affect the abundances of the first low-mass, metal-enriched stars. We find that reverse shock heating can leave the inner mass shells at entropies too high to cool, leading to carbon-enhancement in the re-collapsing gas. This hydrodynamic selection effect could explain the observed incidence of carbon-enhanced metal-poor (CEMP) stars at low metallicity. We further explore how anisotropic ejecta distributions, recently seen in direct numerical simulations of core-collapse explosions, may translate to abundances in metal-poor stars. We find that some of the observed scatter in the Population II abundance ratios can be explained by an incomplete mixing of supernova ejecta, even in the case of only one contributing enrichment event. We demonstrate that the customary hypothesis of fully-mixed ejecta clearly fails if post-explosion hydrodynamics prefers the recycling of some nucleosynthetic products over others. Furthermore, to fully exploit the stellar-archaeological program of constraining the Pop III initial mass function from the observed Pop II abundances, considering these hydrodynamical transport effects is crucial. We discuss applications to the rich chemical structure of ultra-faint dwarf satellite galaxies, to be probed
in unprecedented detail with upcoming spectroscopic surveys.
\end{abstract}


\section{Introduction}
\label{sec:intro}

The first generation of stars, the so-called Population III (Pop III) stars, formed from pure hydrogen, helium, and lithium \citep[for review, see][and references therein]{Bromm:13}. Nucleosynthesis in these stars and their explosions delivered the first metals to enrich the Universe.  Then, second-generation Population II (Pop II) stars formed from sufficiently metal-enriched gas elements.  The first generation of low-mass, long-lived Pop II stars, the relics of which can be found in the Galactic stellar halo and ultra-faint dwarf satellite galaxies (UFDs), encodes a crucial empirical record of Pop III nucleosynthesis \citep{Frebel:15}.  
The principal variable controlling the stellar nucleosynthetic output is the progenitor mass.  Simulations published over the past few years indicate that two processes can limit the masses of Pop III stars: radiative suppression of protostellar accretion \citep{McKee:08,Hosokawa:11,Stacy:12,Hirano:14,Susa:14} and gravitational fragmentation in protostellar disks \citep{Stacy:10,Clark:11a,Clark:11b,Greif:11,Greif:12}.  An upper limit on the production rate of very-low-mass, long-lived Pop III stars can be placed empirically, by surveying for metal-free stars in the Galaxy \citep{Tumlinson:06,Hartwig:15}. The failure to find Pop III stars in the nearby Universe suggests that the characteristic stellar mass was $>0.8M_\odot$. Attempts have been made to speculate about the Pop III initial mass function (IMF).  On the basis of cosmological simulations and still uncertain subgrid prescriptions, \citet{Susa:14} estimated that the stars had masses from $1$ to $300\,M_\odot$, that the IMF peaked at $45M_\odot$, and that there were on average 2.5 Pop III stars per cosmic minihalo.

There are several ways in which the intuition gained in the study of metal-enriched star formation and evolution in the nearby Universe fails in metal-free stars. Pop III stars might have been more rapid rotators, as angular momentum loss through line-driven winds was not efficient in metal-free stellar atmospheres \citep{Stacy:11}.\footnote{It is premature to speculate about differences in magnetically-mediated angular momentum loss because the degree of magnetization of Pop III stars remains uncertain.}  However, rapid rotation drives internal mixing of hydrostatically synthesized elements and this can dredge light elements up into the atmosphere \citep{Ekstrom:08,Yoon:12}. Also, stars lacking metals in their atmospheres have high enough surface temperatures $\sim 10^5\,\mathrm{K}$ to be strong ionizing photon emitters \citep{Schaerer:02}. The associated H\,II regions are initially highly overpressured near the star.  The ionized gas pressure reduces the density of gas into which the Pop III star would explode to $\sim0.1-1\,\mathrm{cm}^{-3}$.  However it has recently become apparent that dense circumstellar gas clumps may avoid photoionization and remain cold and dense until the supernova blastwaves arrive \citep{Abel:07,BlandHawthorn:11,Ritter:12,Ritter:15,Webster:14,Smith:15}.  These clouds modify the mechanics of blastwave-circumstellar medium interaction, for example by securing a reservoir of metal-free gas close to the site in which Pop II stars would eventually form.

The Pop III explosion type and nucleosynthetic yields are sensitive to the mass of the compact remnant produced in the core collapse: the nucleosynthetic products collapsing into or accreted onto the remnant are removed from the yields.  Gross theoretical uncertainties remain, especially with respect to the mass of the black hole produced by fallback and the metal fraction lost in the black hole.  In non-rotating stars, the general expectation is of a core collapse supernova with a neutron star remnant for progenitor masses $10 M_\odot < M < 25 M_\odot$ and a black hole remnant for $25 M_\odot < M < 140 M_\odot$. At still higher masses, a pair instability supernova (PISN)  leaving no compact remnant or a direct collapse into a black hole are the likely outcomes \citep{Nomoto:13}.  Stellar rotation tends to reduce the minimum mass for PISN events \citep{Chatzopoulos:12,Yoon:12}. Estimates of the nucleosynthetic yields of Pop III supernovae as a function of progenitor mass are sensitive to additional theoretically uncertain parameters: the degree of rotation, mixing, and mass loss, the mechanics of fallback (in particular as approximated with a simple mass cut), and explosion energy \citep{Woosley:95,Heger:10,Takahashi:14}. 

Particularly problematic are any conclusions derived from one-dimensional simulations of core-collapse supernovae. Multi-dimensional processes such as convection, instability, and mixing can be included in one-dimensional simulations with sub-grid prescriptions, but the dependence of the amplitude of these effects on progenitor parameters and any potential intrinsic stochasticity are poorly constrained.  For example, explosively-synthesized nickel, the radioactive precursor to iron, can escape gravitational capture in the compact object by getting squirted ahead of the lighter elements in collimated, Rayleigh-Taylor-instability-amplified jets (or ``fingers'').  This, of course, has pivotal implications for chemical enrichment. Under which specific conditions are fingers expected is the subject of recent, rapidly developing research \citep{Hammer:10,Wongwathanarat:13,Wongwathanarat:15,Utrobin:14}.

At a similar stage of maturity as the study of metal-free stars, the narrative of the formation of the first low-mass, metal-enriched stars is still being written. The enhanced cooling that is required to facilitate gravitational fragmentation down to $\lesssim1\,M_\odot$ is provided by dust grains and dust-gas thermal coupling \citep{Omukai:05,Schneider:06,SafranekShrader:14b,SafranekShrader:15}.  Dust is produced in the Pop III supernovae \citep{Schneider:06,Nozawa:07,Cherchneff:10,Chiaki:15} or in red supergiant stellar winds \citep{Nozawa:14}. More speculatively, dust grows in gas phase in the collapse of metal-enriched clouds that would form Pop II stars \citep{Chiaki:14}.  

It has recently become possible to simulate the formation of individual metal enriched stars from first principles, from the initial conditions imprinted in the Big Bang and by tracking all the relevant dynamical, chemical, and radiative processes  \citep{SafranekShrader:14b,SafranekShrader:15}.   These challenging simulations have not yet converged to making a definite prediction for, say, the shape of the Pop II stellar IMF, even in very-high-redshift systems in which the separation of the halo virial and star-forming time scales is the smallest and least computationally prohibitive.  We are still some ways from a predictive theory of star formation across the halo mass scales and redshifts at which first-generation metal-enriched star formation is expected to take place.

All these theoretical bottlenecks motivate the aspiration that the nucleosynthetic yields of Pop III stars can be reverse-engineered from chemical abundances measured \citep[e.g.,][]{Cayrel:04} in low-mass, long-lived stars in what can be called ``stellar archaeology''.  Because the most metal-poor stars may have formed from gas enriched by one or a small number of Pop III supernovae, measured stellar chemical abundances can be used to test theories of Pop III nucleosynthesis \citep{Frebel:05,Frebel:15}. Theoretical predictions of isotopic yields are sensitive to parameters not currently determined from first principles. Therefore the alternative approach of using the observed abundance ratios in the most metal-poor stars to identify classes of contributing supernovae in the chemical abundance space may be more promising \citep{Ting:12,Tominaga:14}. For instance, \citet{Ishigaki:14} have shown that the most iron-poor stars can be explained as having been enriched by low-energy Pop III core-collapse explosions in which most of the iron was lost in the compact remnant. Indeed, faint supernovae with significant fallback seem to be necessary to explain the abundances of the most metal-poor stars \citep{Iwamoto:05}. The hypothetical ultra-energetic, magnetic-outflow-powered ``hypernovae'' may explain  abundance ratios in extremely metal poor stars \citep{Izutani:10}. The effects of rotation in Pop III stars may affect the C, N, and O abundances, but despite the expectation that Pop III stars were rapid rotators, very few nucleosynthesis calculations take rotation into account \citep{Ekstrom:08,Joggerst:09,Joggerst:10}.

Metal-poor stars that could have been enriched by Pop III stars exist in the inner and outer stellar halo of the Milky Way, in old, metal-poor globular clusters, and in dwarf satellite galaxies.  Especially promising for stellar archaeology are the UFDs \citep{Kirby:08,Norris:10b,Simon:10,Brown:12,Vargas:13,Frebel:14}. The inner and outer Galactic halo have peak metallicities of $[\mbox{Fe/H}]=-1.6$ and $-2.2$, respectively \citep{York:00,Gunn:06,Carollo:08}.\footnote{We employ the standard notation: $[A/B]=\log(A/B)-\log(A_\odot/B_\odot)$ where $A$ and $B$ are mass-fractions of isotopes $A$ and $B$ while $A_\odot$ and $B_\odot$ are the corresponding solar mass fractions.} The fraction of stars that are carbon enhanced and yet metal poor (CEMP, defined as $[\mbox{C/Fe}]>+1.0$ and $[\mbox{Fe/H}]<-1.0$) increases as the metallicity decreases; thus the outer halo has a higher CEMP fraction \citep{Yong:13}. Stars in UFDs have metallicity and abundance ratio distributions similar to those of the halo stars but with a few key differences \citep{Corlies:13}. The abundances of neutron capture elements in UFDs are lower than in halo stars \citep{Frebel:10b} while $\alpha$-element abundances are more enhanced in UFDs \citep{Kirby:11}.

In addition to metal poor stars in the local group, the earliest stages of chemical evolution can be studied in intergalactic absorption line systems such as the damped Lyman-$\alpha$ (DLA) systems on quasar \citep{Cooke:11a,Cooke:11b,Cooke:15,Becker:12} and gamma-ray burst \citep{Lamb:00,Wang:12,Cucchiara:15,Ma:15} sightlines. There has been an attempt to constrain the Pop III IMF by measuring the relative abundances of various elements in a sample of $\sim10$ DLAs \citep{Kulkarni:13}. By comparing the abundance ratios in DLAs with those of halo and dwarf galaxy stars, we can learn how the systems that polluted the DLAs are related to the relic stellar systems in the nearby Universe \citep{Salvadori:12,Webster:15}.   

Various explanations have been offered for the observed abundance patterns in metal poor stars and, in particular, the metallicity distribution function and the scatter in abundance ratios as a function of [Fe/H]. Stars with the lowest metallicities may have formed from gas enriched by only one or possibly a few Pop III supernovae \citep{Audouze:95,Feltzing:09}. The halo-to-halo variation of Pop III properties \citep{Hirano:14} could explain the relatively large scatter in abundance ratios that has been observed in the lowest metallicity stars \citep{Frebel:12}. The detailed abundances can be explained if most of the Pop III stars produced core collapse supernovae and only a few exploded as PISNe \citep{Aoki:14}.

The decrease of the carbon-enhanced fraction with metallicity can be explained if more massive Pop III stars produce PISNe that synthesize more iron but also remove more gas from the halo (through photoionization and supernova blasting), thus making it less likely that iron-rich Pop II stars will subsequently form in that halo \citep{Cooke:14}. The existence of carbon-normal and carbon-enhanced metal-poor stars has also been interpreted as evidence for two distinct cooling channels in early metal-enriched star forming systems. In this scenario, CEMP stars form from gas that is rich in C and O and first cools via atomic fine structure transitions, while carbon-normal stars form via non-carbon (e.g., silicate) dust-mediated cooling \citep{Norris:13,Ji:14}. The decrease in abundance scatter with metallicity is then a consequence of an increase of the number of contributing nucleosynthetic events \citep{Argast:00}.  However it remains unclear if the hypothetical dust-free, C- and O-enriched gas can indeed efficiently fragment into low-mass protostellar cores.

This is our third paper in a sequence \citep{Ritter:12,Ritter:15} investigating the small-scale post-supernova evolution affecting interpretations of this record.
Recognizing that the character of the Pop III-to-II transition can be sensitive to the poorly-explored small-scale hydrodynamic effects in supernova remnant (SNR) evolution, \citet{Ritter:12} followed the metal-enriched ejecta from a single core-collapse-type Pop III explosion in a $10^6 M_\odot$ minihalo---starting from the early, kinetic-energy dominated free expansion phase---for about $40\u{Myr}$ after the explosion. They found that the metals were inhomogeneously dispersed in the host halo and its immediate neighborhood. The metal-enriched gas returning into the halo center became diluted with primordial gas streaming from the cosmic web only in a dense, numerically-unresolved gaseous core at the center of the simulation. 

Proceeding to a higher spatial resolution, \citet{Ritter:15} showed that turbulence excited by gravitational infall in a collapsing gaseous core is sufficient to homogenize chemical abundances in the highest-density gas.  \citet{Feng:14} showed that turbulent homogenization happens in collapsing, star-cluster-forming clouds in the present-day Galactic disk \citep[see, also,][]{BlandHawthorn:10}. Interestingly, however, \citet{Ritter:15} also found that the high-density homogenization did not uniformly sample the abundance profiles initially injected by supernovae. There seemed to be a trend in which the yields that were raised to higher entropies in blastwave shocks were deficient in the fallback and recycling into subsequent stellar generations.  Ritter et al.\ suggested that this trend arose from a positive correlation between the cooling time in the supernova-ejecta-enriched gas and the specific entropy to which the reverse supernova shock has raised the ejecta that would enrich the gas.

Here, we continue analysis of one of the two simulations of \citet{Ritter:15} to find that abundance ratios in the ejecta of a Pop III supernova need not be preserved in star-forming clouds.  A non-trivial, stochastic relation between yields and enrichment has to be taken into account when comparing the abundances measured in metal-poor stars with the predicted yields of supernovae. The mismatch arises from a combination of two hydrodynamical effects, one related to the realization that the yields can be anisotropically distributed in the ejecta, and the other that the reverse shock raises the inner ejecta to entropies too high to cool and be recycled into new stars.  Whereas \citet{Ritter:15} focused on the case of multiple supernovae,
  we here treat the case of inhomogeneous and anisotropic mixing biases
  from a single Pop III explosion. We thus derive, as it were, the 
  irreducible degree of such biases, and investigate how they translate
  into a minimal degree of scatter in the empirical abundance record.

The rest of the paper is organized as follows.  In Section \ref{sec:method} we describe our numerical methodology, including in particular detail left out in \citet{Ritter:15}.  In Section \ref{sec:results} we discuss morphological evolution of the SNR, ejecta dispersal trends, chemical variation in diffuse and dense gas, and explore implications for identifying chemical abundance biases and anomalies in observations of metal-poor stars.  In Section \ref{sec:discussion_conclusions} we briefly reflect on the broader state of stellar archaeology. 

\section{Methodology}
\label{sec:method}

\subsection{Initial conditions, dark matter, and hydrodynamics}


The initial conditions were mapped from a snapshot of the gravitational-hydrodynamic simulation reported in \citet{Ritter:12}. Briefly, the parent simulation had been carried out in a periodic cosmological box of comoving volume $1\u{Mpc}^3$.  Gaussian perturbations in the parent simulation were initialized at redshift of $z=146$ with the code \textsc{graphic2} of \citet{Bertschinger:01}.  Chemical composition of the gas was evolved with a non-equilibrium chemical network for the primordial chemical species as in \citet{SafranekShrader:12}.  At redshift $z=19.7$, a $10^6\,M_\odot$ virialized halo formed and near the center of the halo the gas density reached $n\sim 10^3\u{cm}^{-3}$.  At this time, we inserted a star particle at the density maximum.  The star particle accelerated under the combined gravitational force of the gas and dark matter but was otherwise collisionless.

To initialize the present simulation, we took the cosmological simulation at the instance of star particle insertion and excised a cubical region of physical width $1\u{kpc}$ centered on the particle. Hydrodynamic variables in the region were mapped onto an isolated, non-expanding cubical box of the same size as the excised region.  The center-of-mass velocity in the mapped domain was subtracted from the gas velocity in each cell. The density, temperature, and pressure were mapped directly from the parent grid. We did not map the dark matter particles.  Instead, we modeled the dark matter mass distribution 
with a time-dependent power-law density profile that was spherically symmetric around the box center. From the insertion of the star particle (the beginning of the simulation) until
 the insertion of the SNR, the dark matter density profile was held constant
\beq
\label{eq:rho_constant}
\rho_{\rm DM}(r) = 940\,r_{\rm pc}^{-2.132}\,M_\odot\,\mathrm{pc}^{-3}
\eeq
where $r_{\rm pc}$ is the distance from the center of the halo in parsecs. This profile was obtained as a direct spherically-symmetric power-law fit to the dark matter density in the parent simulation at the point of star particle insertion.  To model dark matter halo growth from $1\,\u{Myr}$ after SNR insertion onward, we implemented a time dependence through a scaling calibrated against the halo growth rate in the \citet{Ritter:15} continuation of the simulation over $200\,\u{Myr}$ after supernova insertion
\bea
\label{eq:rho_evolving}
\rho_{\rm DM}(r,t) = f(t)\, g(t)^{-3} \rho_{\rm DM}(g(t)^{-1} r) ,
\eea
where $f(t)=1.029^{t_{\rm Myr}}$, $g(t)=1.012^{t_{\rm Myr}}$, and $t_{\rm Myr}$ is time after supernova insertion in Myr.

The simulation was carried out using the adaptive mesh refinement (AMR) hydrodynamics code \textsc{flash}, version 4.0 \citep{Fryxell:00,Dubey:08}.  We used the directionally-split piecewise-parabolic method in three dimensions to solve the hydrodynamics equations.  The gravitational potential was computed by adding the analytical dark matter density profile in Equations (\ref{eq:rho_constant}) and (\ref{eq:rho_evolving}) to the gas density on the AMR grid and then solving the Poisson equation with the \textsc{flash} multigrid iteration scheme \citep{Ricker:08} with isolated boundary conditions.  The simulation was not cosmological in the sense that it did not include universal expansion and active dark matter particles.  The size of the mapped box was just large enough to avoid corruption of the solution at the box center by boundary artifacts over the $60\,\u{Myr}$ of evolution.


\begin{figure*}
\begin{center}
\includegraphics[trim=0cm 1.5cm 0cm 6cm,clip=true,width=0.49\textwidth]{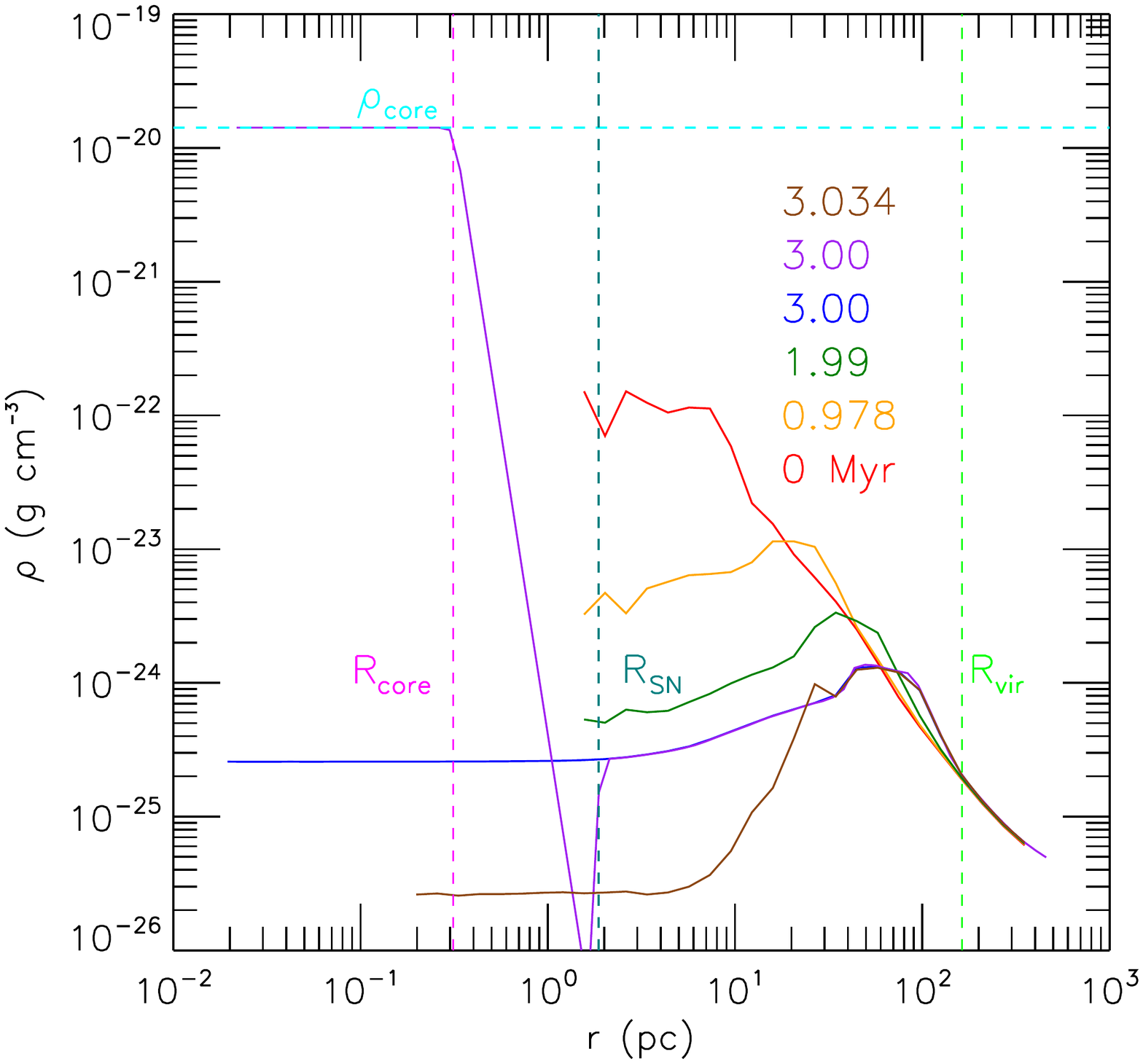}
\includegraphics[trim=0cm 1.5cm 0cm 6cm,clip=true,width=0.49\textwidth]{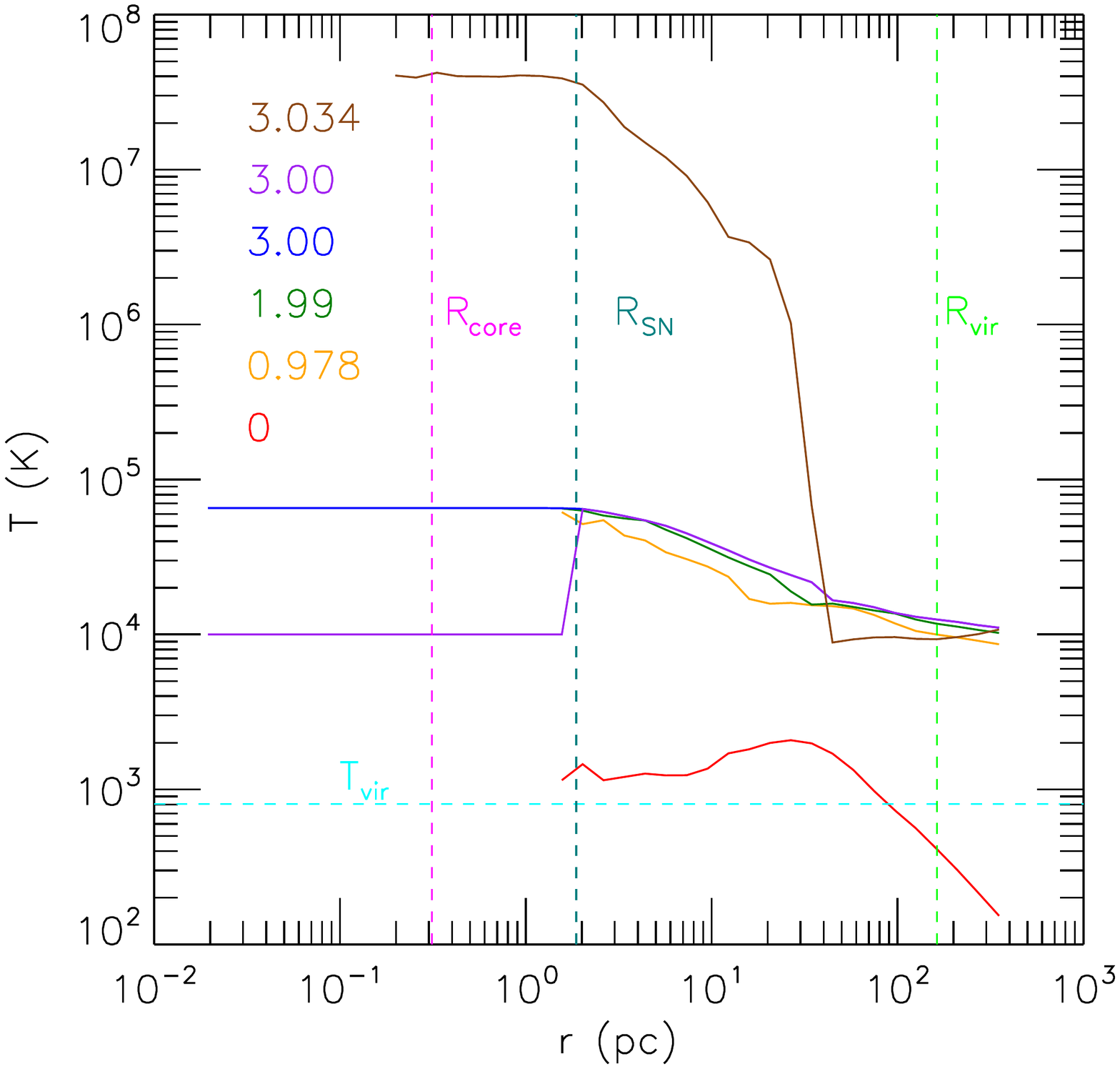}
\end{center}
\caption{Radial profiles of density and temperature at four times before and two times after SNR insertion. The profiles were spherically averaged around the star particle before insertion and around the insertion site after insertion. The time since the beginning of the simulation is indicated in the legend. The blue line labeled $3\u{Myr}$ shows the time immediately before SNR insertion and the purple line also labeled $3\u{Myr}$ shows the time immediately after SNR insertion. The brown line labeled $3.034\u{Myr}$ shows the time when the reverse shock has first reached the center of the halo. Also shown as dotted lines are the initial core density $\rho_{\rm core}$, the core radius $R_{\rm core}$, and the initial SNR radius $R_{\rm SN}$, as well as the virial radius $R_{\rm vir}$ and temperature $T_{\rm vir}$ of the halo.}
\label{fig:radialHII}
\end{figure*}

\subsection{Pop III stellar model, photoionization, and cooling}

We used a single collisionless star particle to represent the position of the Pop III star cluster that we assumed had formed at the center of the halo. The star particle acted as an isotropic source of ionizing radiation with luminosity $Q(\mbox{H})=6\times 10^{49}\u{ionizing photons}\u{s}^{-1}$ over a stellar lifetime of $3\u{Myr}$.   The particle was placed at the gas density maximum at the beginning of the simulation.  For reference, if $Q(\mathrm{H})$ was to be emitted by single metal-free star, its mass would have been $\sim70\,M_\odot$ \citep{Schaerer:02}.  We consistently set our ejecta mass to be lower than this, $40\,M_\odot$, to allow for mass loss to a compact remnant and for additional, longer-living Pop III stars that could have formed in the primary star's protostellar disk and contributed their own ionizing radiation (see Section \ref{sec:SNR}).

We assumed photoionization equilibrium in cells exposed to ionizing radiation from the star and collisional ionization equilibrium in all other cells. For photoionization equilibrium we used the code \textsc{cloudy} \citep{Ferland:13} to tabulate the temperature and ionization state as a function of the ionization parameter, $\xi \sim F/n$, where $F$ is the ionizing photon flux and $n$ is the gas number density.   For collisional ionization equilibrium, we used  \textsc{cloudy}  to tabulate the ionization state and cooling rate as a function of density, temperature, and metallicity. The tabulated data was interpolated at run time.  Cooling was operator-split from the hydrodynamic update.   Where necessary, cooling was sub-cycled on time steps equal to one-tenth of the local cooling time.  

As a simplification, we did not include molecules and dust grains in the equilibrium calculations.The tracking of molecule formation would have required the integration of a non-equilibrium chemical rate equation, something we did in the parent cosmological simulation but omitted here for the sake of computational efficiency.   The absence of H$_2$ molecules  in the recollapsing remnant might have artificially delayed gas collapse past densities $\sim10\,\mathrm{cm}^{-3}$.  Had H$_2$ or dust been present, the runaway gravitational collapse could have potentially set in much earlier.

The details of ionizing radiative transfer were the same as in \citet{Ritter:15}. Briefly, the cartesian AMR grid was mapped onto a spherical grid defined with the HEALPix scheme \citep{Gorski:05}. Along each of the $\sim3000$ angular pixels and out to each pixel-specific Str\"omgren radius, we transported the ionizing flux between bins and then mapped it back onto the AMR grid. In the \textsc{cloudy} calculations, the source was assumed to have a Planck spectral energy distribution at  $T_{\rm eff} = 10^5 \u{K}$.  The method is not precisely photon-conserving but was sufficient to reproduce the correct hydrodynamic response of the gas to photoionization heating (radiation pressure was not computed).
Figure~\ref{fig:radialHII} shows the effect of ionization on the density and temperature of the gas in the halo. When the star particle is inserted, the gas density at the center of the halo is $\sim 60\,\u{cm}^{-3}$ and the temperature is $\sim 1000\u{K}$. The temperature quickly rises to a central maximum of $\approx 70,000\u{K}$ and the pressure excess drives the density to gradually drop to $\sim  0.2\u{cm}^{-3}$ within the central few parsecs after $3\,\u{Myr}$.

\begin{figure*}
\begin{center}
\includegraphics[width=0.33\textwidth]{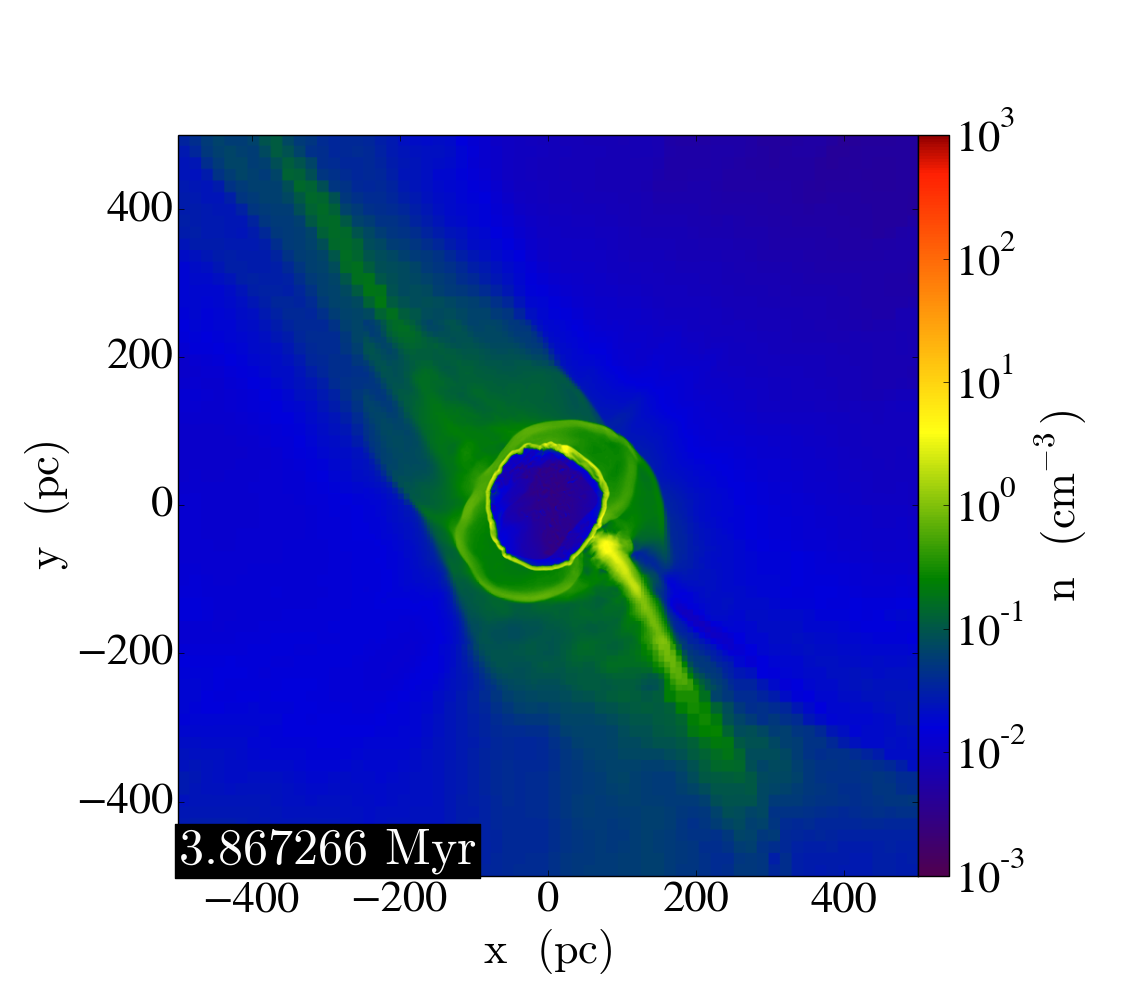}
\includegraphics[width=0.33\textwidth]{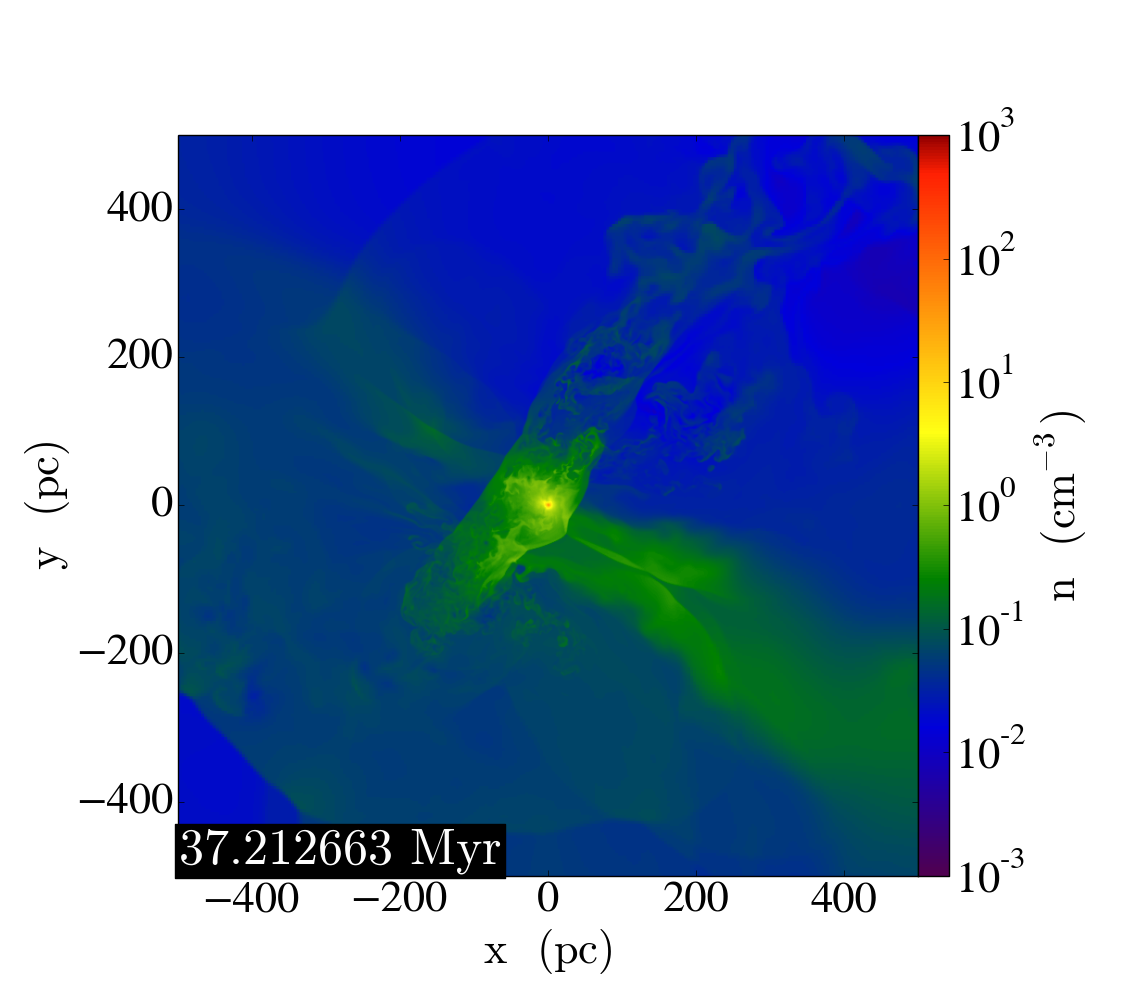}
\includegraphics[width=0.33\textwidth]{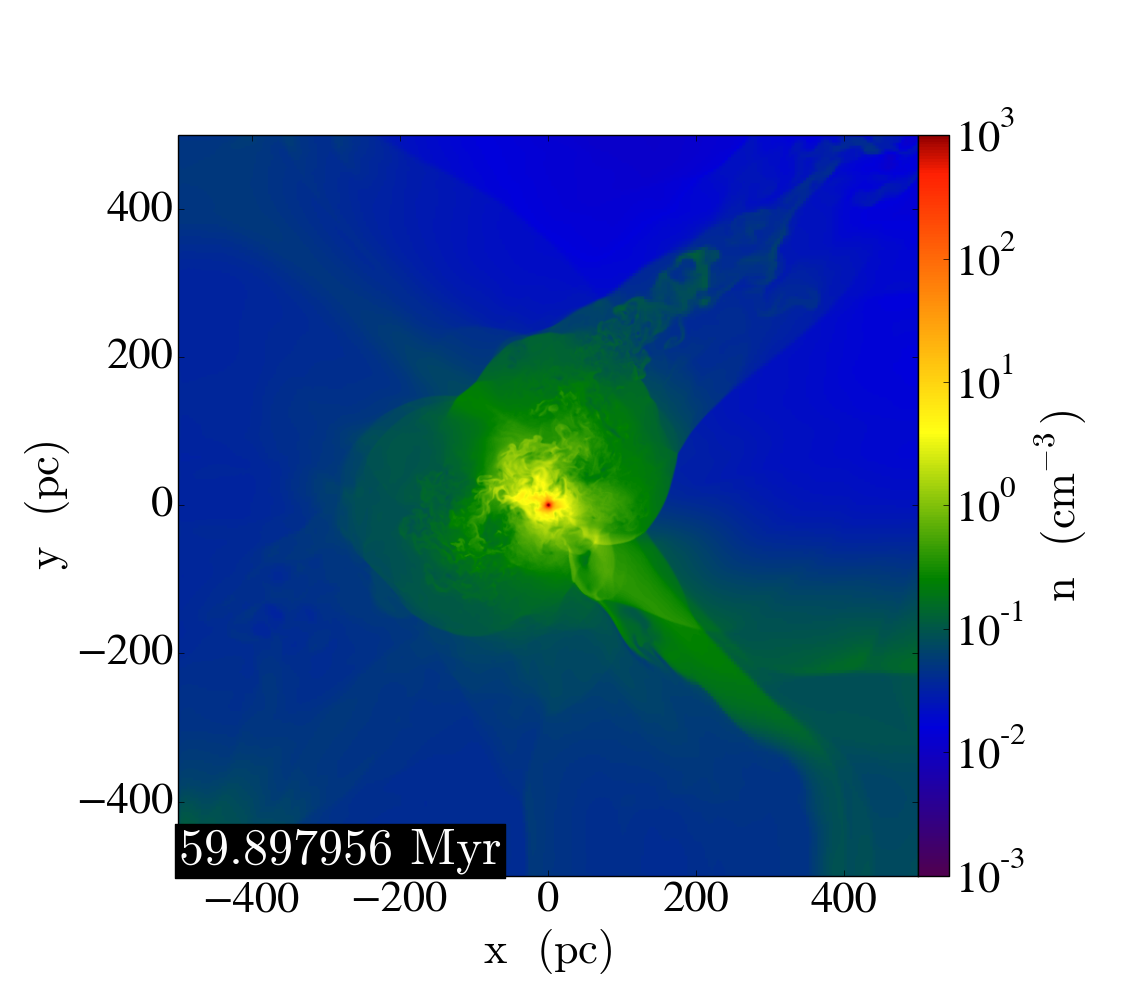}
\includegraphics[width=0.33\textwidth]{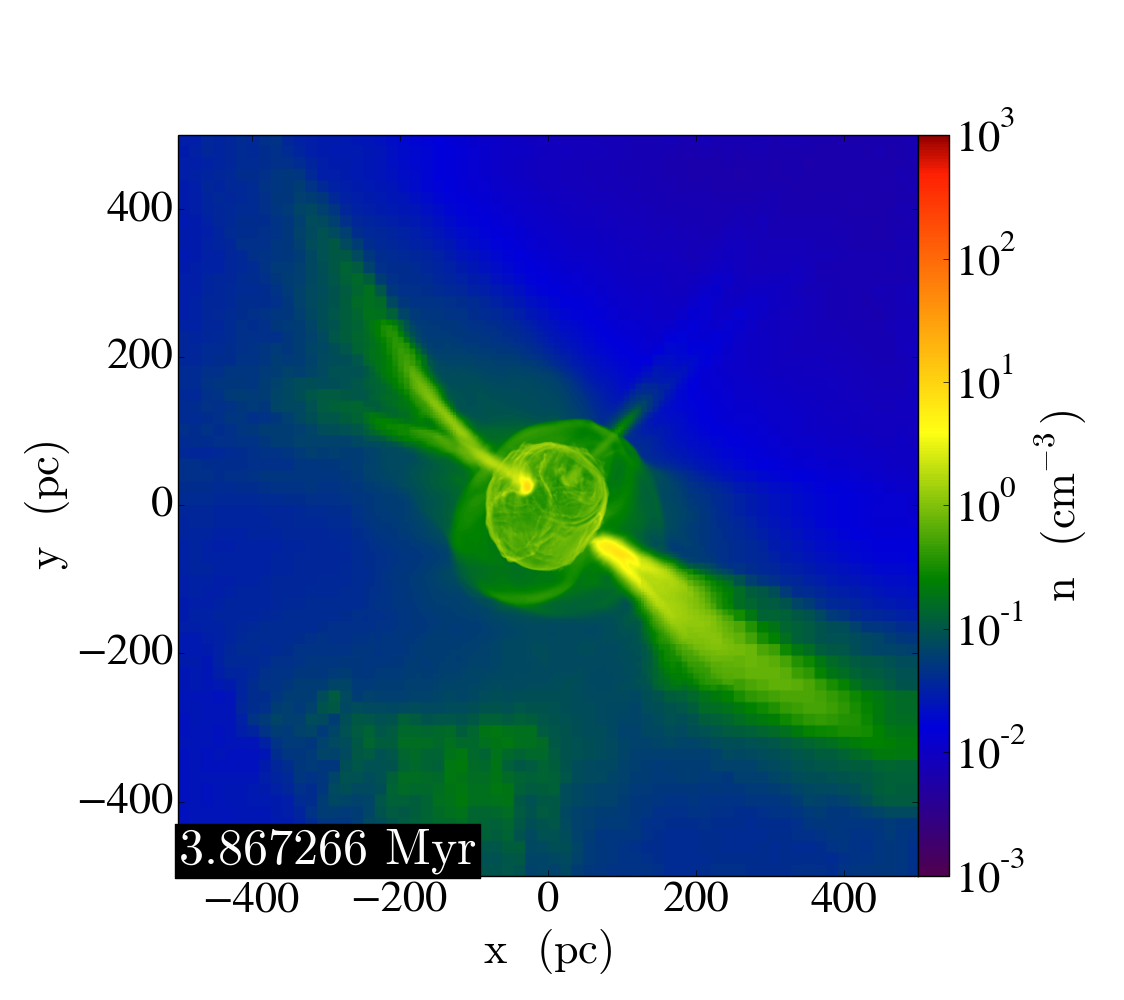}
\includegraphics[width=0.33\textwidth]{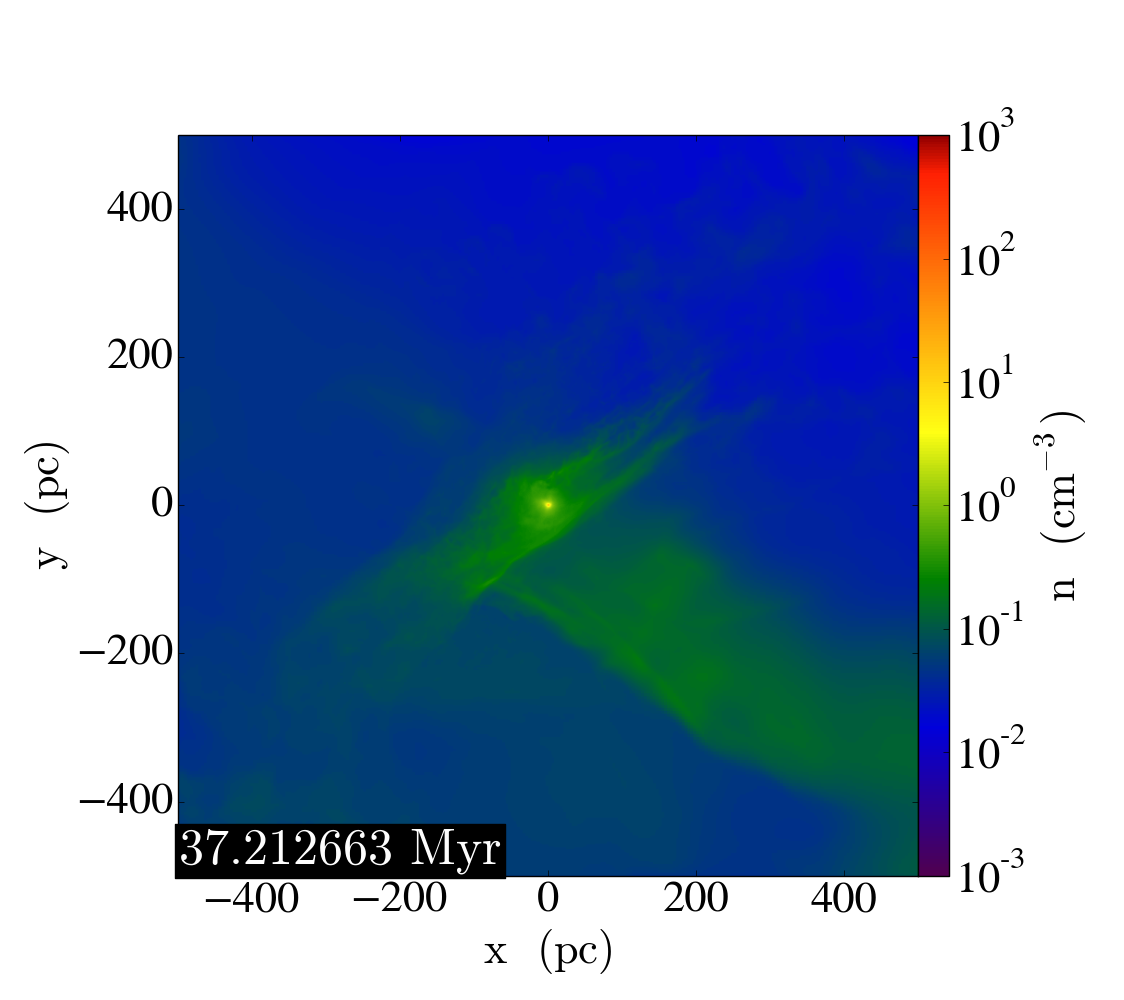}
\includegraphics[width=0.33\textwidth]{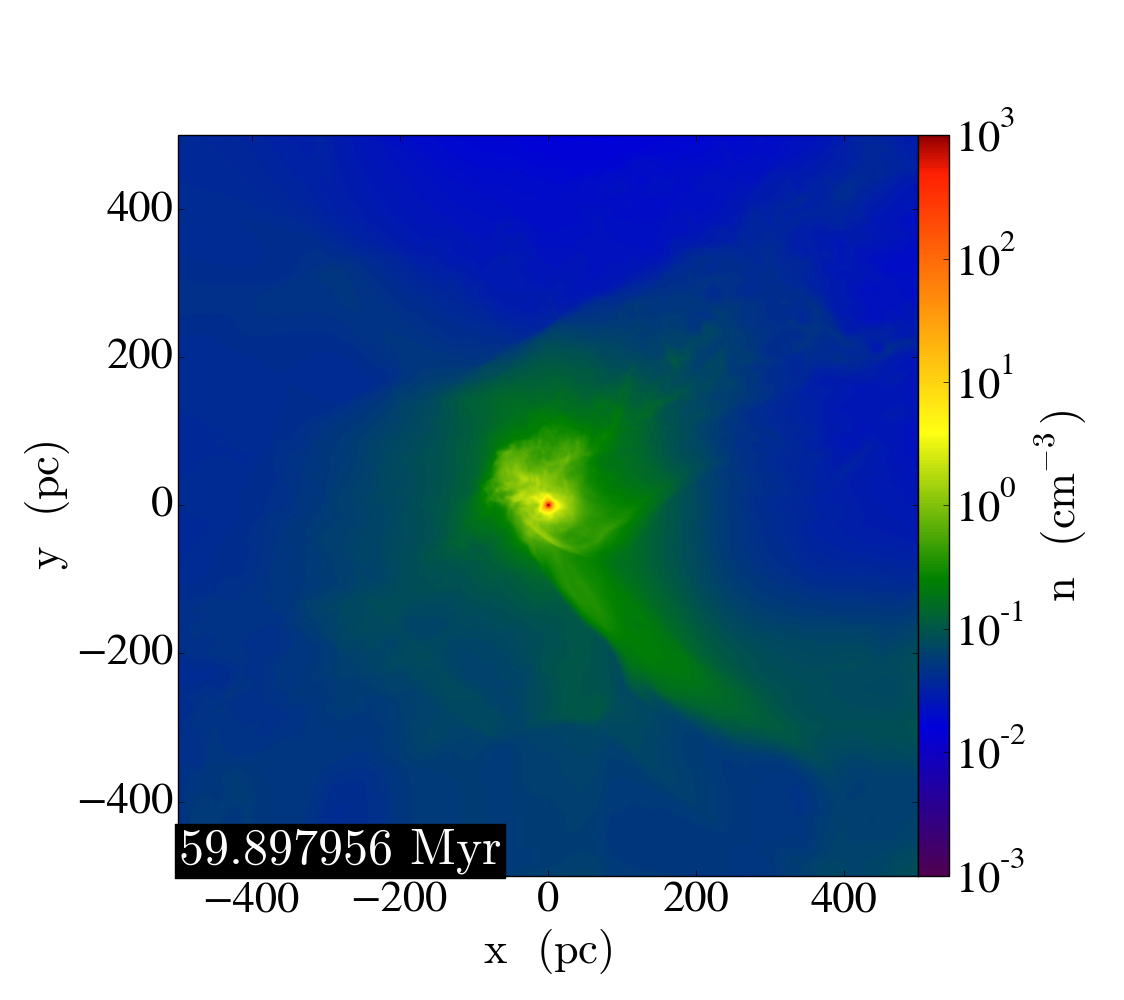}
\includegraphics[width=0.33\textwidth]{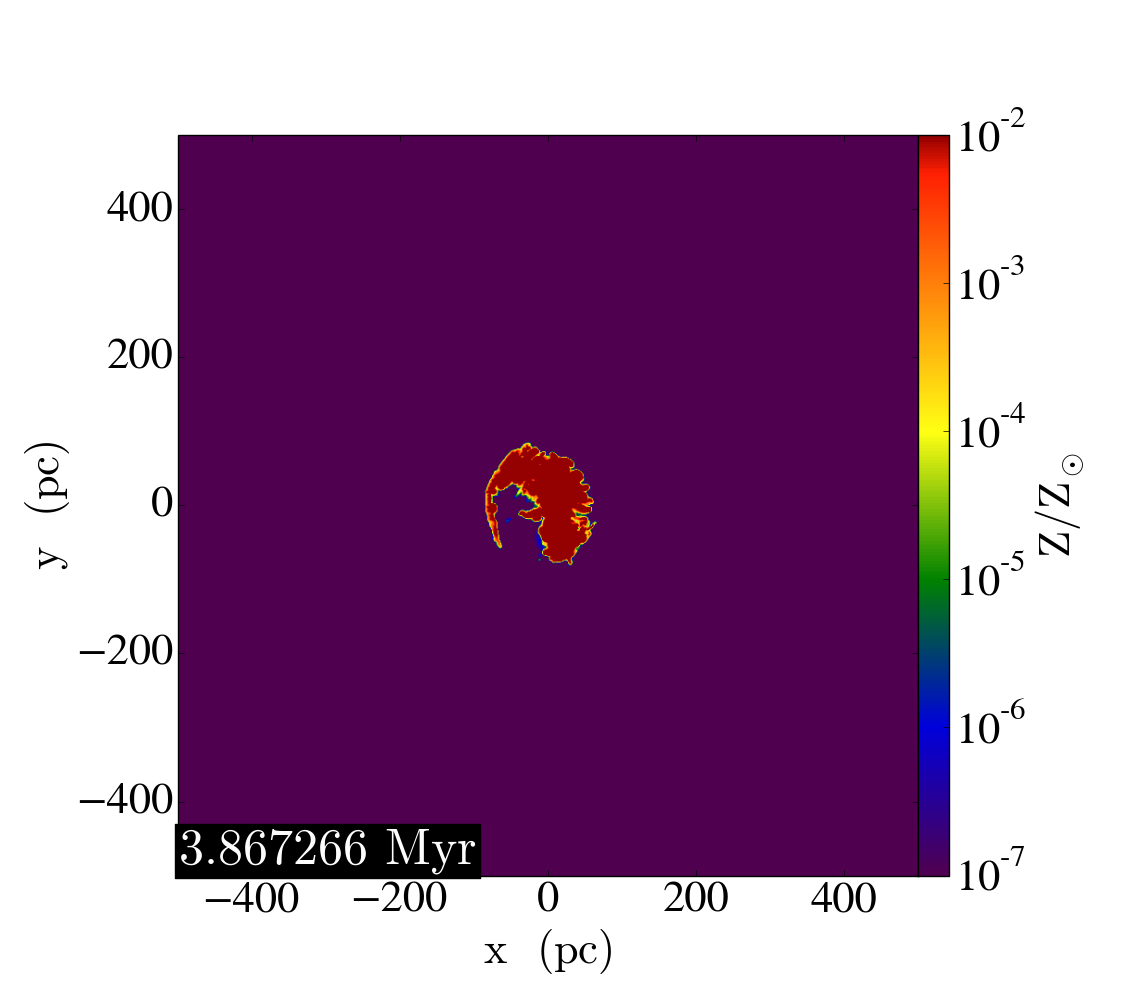}
\includegraphics[width=0.33\textwidth]{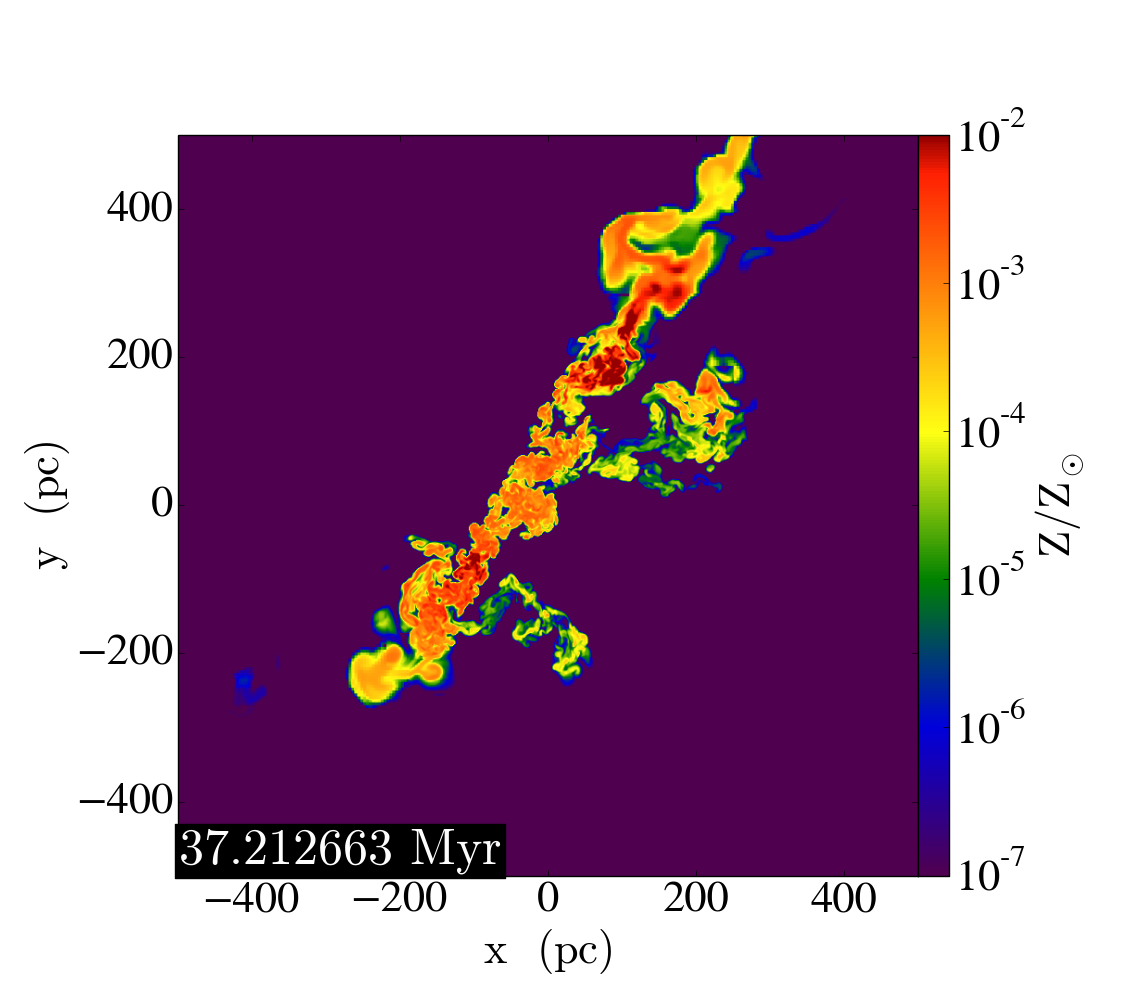}
\includegraphics[width=0.33\textwidth]{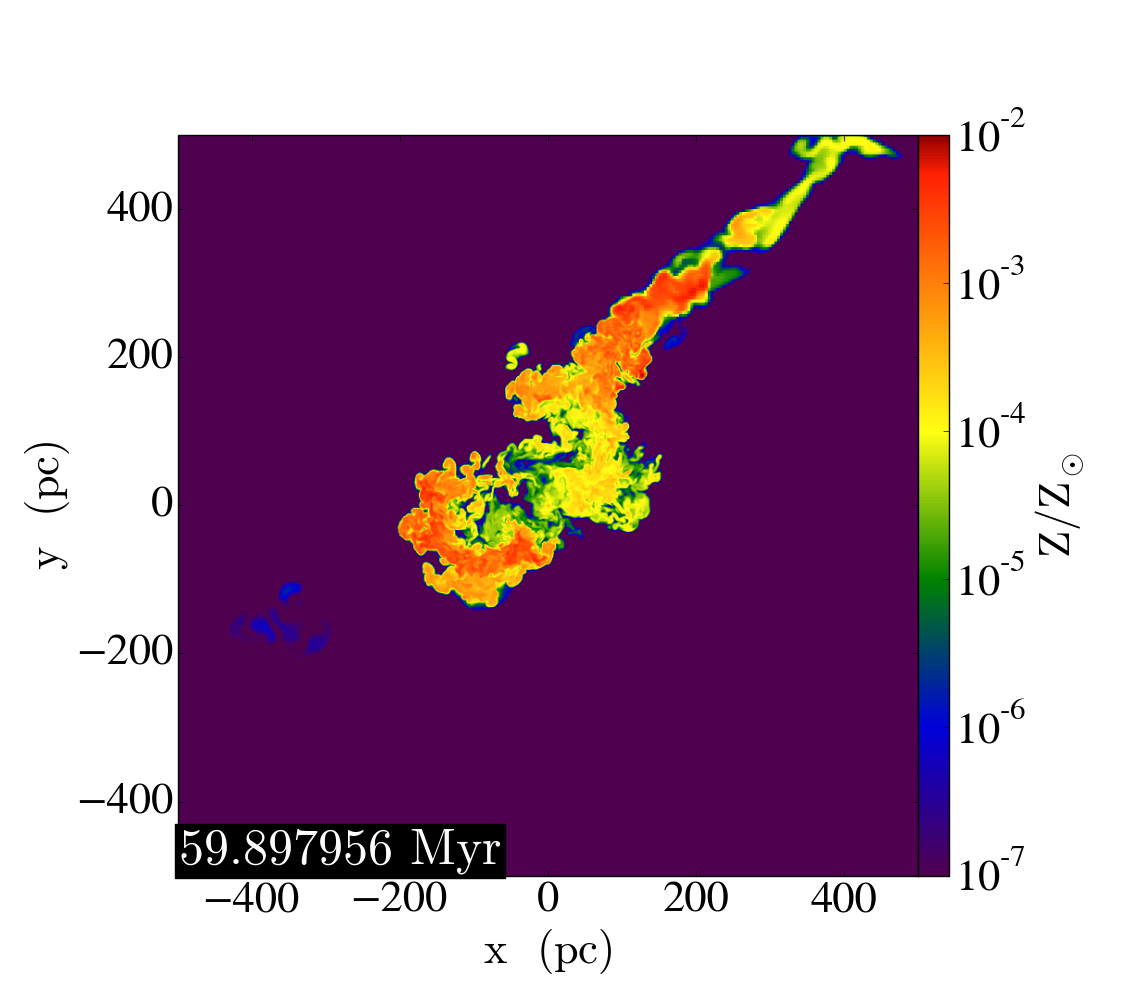}
\includegraphics[width=0.33\textwidth]{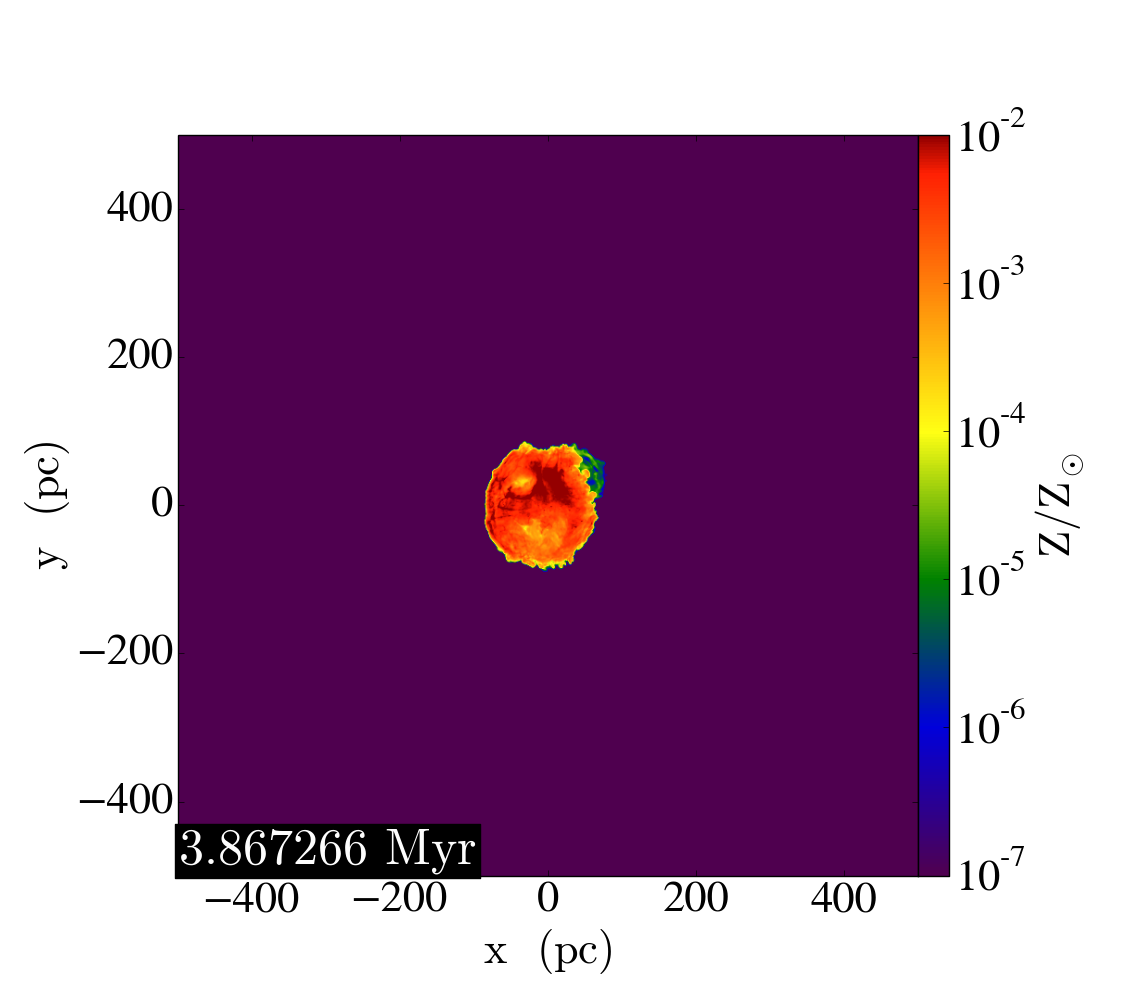}
\includegraphics[width=0.33\textwidth]{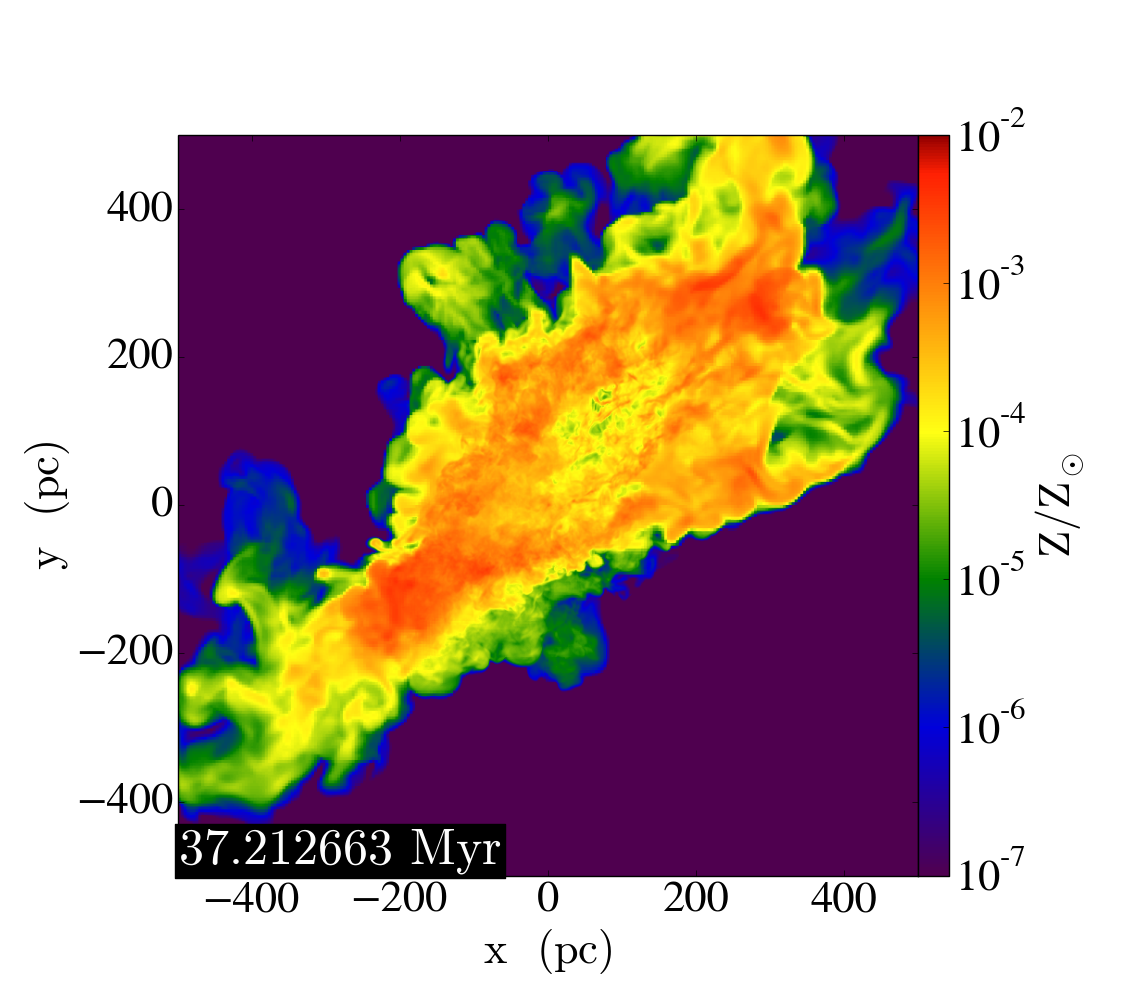}
\includegraphics[width=0.33\textwidth]{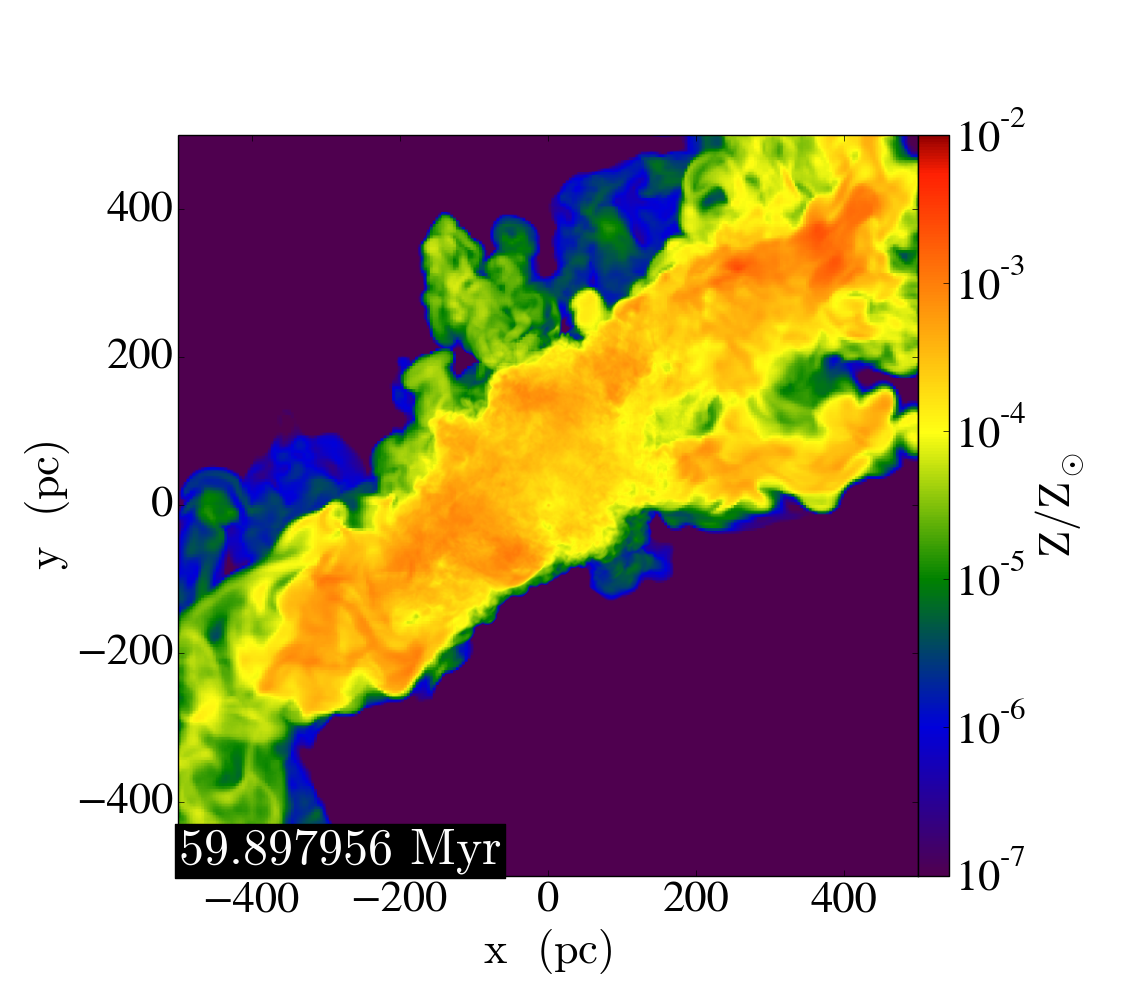}
\end{center}
\caption{
The columns show the evolution of the remnant at three different times after insertion, where the time since the beginning of the simulation is shown in the bottom left corner of each plot (the SNR was inserted at $3\u{Myr}$). 
The rows show: density slices (top row), density-weighted gas density projections ($\int \rho n dz/\int\rho dz$; second row), passive scalar metallicity slices (third row), and density-weighted passive scalar metallicity projections ($\int \rho Zdz/\int \rho dz$; bottom row).  Note that the metallicity color scale is truncated to display metallicities exceeding $0.01\,Z_\odot$ with the same color even where the local metallicity substantially exceeds this.}
\label{fig:slice_projection}
\end{figure*}

\subsection{Supernova remnant and ejecta tracking}
\label{sec:SNR}

The ionizing source was turned off $3\u{Myr}$ after star particle insertion and a SNR was inserted at the location of the star particle. The initial radius of the remnant of $R_{\rm SN}=1.87\u{pc}$ was chosen such that the remnant was in free expansion and the blastwave has not had the chance to sweep a circumstellar gas mass comparable to the ejecta mass.  The SNR was inserted by modifying the gas density, temperature, absolute metallicity, and gas velocity in a sphere of radius $R_{\rm SN}=1.87\u{pc}$.  We followed the prescription of \citet{Whalen:08} to set the density and velocity to
\bea
\rho \= \left\{
\begin{array}{ll}
\rho_{\rm core} , & r<R_{\rm core} , \\
\rho_{\rm core}  \left(\frac{|{\mathbf r}-{\mathbf r}_{\rm SN}|}{R_{\rm core}}\right)^{-\beta} , & r\ge R_{\rm core} ,
\end{array}
\right.
\eea
and
\bea
\bold v \= \f{v_{\rm max}}{R_{\rm SN}} (\bold r - \bold r_{\rm SN}) ,
\eea
where $\mathbf{r}_{\rm SN}$ denotes the center of the SNR, $v_{\rm max}=10^9 \u{cm}\u{s}^{-1}$, and $\beta=9$.  The SNR core density $\rho_{\rm core}$ was computed with
\bea
\rho_{\rm core} \= \f{3 M_{\rm ej}}{4\pi R_{\rm SN}^3} 
\({1-\f{\beta}{3}}\nonumber\\ & & \times
\[{1-\f{\beta}{3}\({\f{R_{\rm core}}{R_{\rm SN}}}^{3-\beta}}^{-1}
\({\f{R_{\rm core}}{R_{\rm SN}}}^{-\beta}\nonumber\\
&=& 1.42 \times 10^{-20} \u{g}\u{cm}^{-3}
\eea
and the core radius with
\bea
R_{\rm core} &=& \f{R_{\rm SN}}{v_{\rm max}} \sqrt{\f{10(\beta-5)E_{\rm kin}}{3 (\beta-3)M_{\rm ej}}} \nonumber\\
&=& 9.62 \times 10^{17} \u{cm} ,
\eea
where $E_{\rm kin}=10^{51}\u{erg}$ is the kinetic energy of the explosion and $M_{\rm ej}=40\,M_\odot$ is the ejecta mass.  The temperature was set to an arbitrary uniform initial value of $10^4\u{K}$.  Since we do not simulate molecule and dust formation, the temperature of the ejecta preceding reverse shock passage, as long as it is low enough ($\ll 10^7\u{K}$) to ensure that kinetic energy dominates, is irrelevant for the physical outcome of the simulation.

The instance of SNR insertion corresponds to a physical time of $R_{\rm SN}/v_{\rm max}=182\u{yr}$ after core-collapse explosion. The total pre-explosion gas mass displaced by SNR insertion was $0.08\,M_\odot$, negligible compared to the ejecta mass.  The small displaced mass justifies our neglect of the blast wave-circumstellar matter interaction in the period preceding insertion.













We track the metal-bearing ejecta redundantly in two ways: with an Eulerian passive scalar variable representing the absolute metallicity, and with a large number of Lagrangian tracer particles with initial distribution matching the ejecta density. Redundant tracking is required to control the artificial numerical diffusion to which Eulerian passive scalar advection is susceptible. This diffusion introduces spurious quasi-exponential low metallicity tails where metal-free gas is in contact with metal rich gas \citep[see][for further discussion of numerical diffusion in passive scalar transport]{Ritter:12,Ritter:15}.\footnote{\citet{Obergaulinger:14} have recently employed the tracer particle technique to study ejecta hydrodynamics in the Vela Jr.\ SNR.}

We assumed that $4\,M_\odot$ of the ejecta was in metals.  We compute the cooling rate and ionization state of the gas from the local passive scalar metallicity which for simplicity we treated as fully mixed within the initial remnant so that metallicity within the remnant was $Z=0.1\approx 5\,Z_\odot$ and abundances were $\alpha$-enhanced ($[\alpha/\mathrm{Fe}]=0.5$) but otherwise solar (in contrast with the metal transport analysis in Section \ref{sec:results} where mixing and $\alpha$-enhanced solar abundances were not assumed).

At $1\u{kyr}$ after the SNR was inserted in the simulation, $10^7$ passive tracer particles were also inserted to track the motion of the ejecta.  The particles were advected with the gas velocity cubically interpolated from the computational grid. Each particle represented a mass of $M_{\rm part} =4\times 10^{-6\,}M_\odot$ of ejecta gas.  The number of particles placed in each AMR cell of volume $\Delta V$ was set to a random integer having the same mean as, and within unity of $Z\, \rho\, \Delta V/M_{\rm part}$, where $Z$ and $\rho$ are the passive scalar metallicity and density in the cell.

Each Lagrangian tracer particle was given a unique tag so that the particle's location can be traced for the duration of the simulation.  A spherical coordinate system $(r,\theta,\phi)$ was centered on the explosion with $\theta$ being the inclination with respect to the positive $z$-axis and $\phi$ the azimuth relative to the positive $x$-axis. The particles were then classified by these coordinates. At the end of the simulation particles that were located in cells with gas densities exceeding $10^{-21}\u{g}\u{cm}^{-3}$ were treated as having the chance to having been incorporated in a Pop II star cluster that would form at the center of the halo if the hydrodynamical evolution were tracked to higher densities and spatial resolutions.  Particle tagging allowed us to trace every particle that made it into the Pop II cluster to its initial location in the supernova ejecta. 

\subsection{Adaptive refinement}

While the ionizing source was present in the simulation, the computational block ($8\times8\times8$ computational cells) containing the source was forced to be at refinement level $\ell_{\rm max}=7$ and the cell size was $\Delta x_{\rm min} = 1\u{kpc}/2^{\ell_{\rm max}+2}\approx 2\u{pc}$. The other blocks were at the lowest refinement level possible subject to the \textsc{flash} constraint that adjacent computational blocks can differ in refinement level by at most one ($|\Delta \ell |\leq 1$).
At the end of the star's life ($t_{\rm SN}=3\u{Myr}$), the hydrodynamic time step was reset to a small value, the ionizing source was turned off, and the block containing the star particle was refined so that the particle now found itself in a block at a maximum refinement level of $\ell_{\rm max}=13$ with cell size $\Delta x_{\rm min} \approx 0.03\u{pc}$. The region at maximum refinement was made just large enough to let the entire SNR be inserted at the same, uniform refinement level. As before, the blocks not required to be at the maximum refinement level were derefined as much as AMR constraints allowed.

\begin{figure}
\begin{center}
\includegraphics[width=0.48\textwidth]{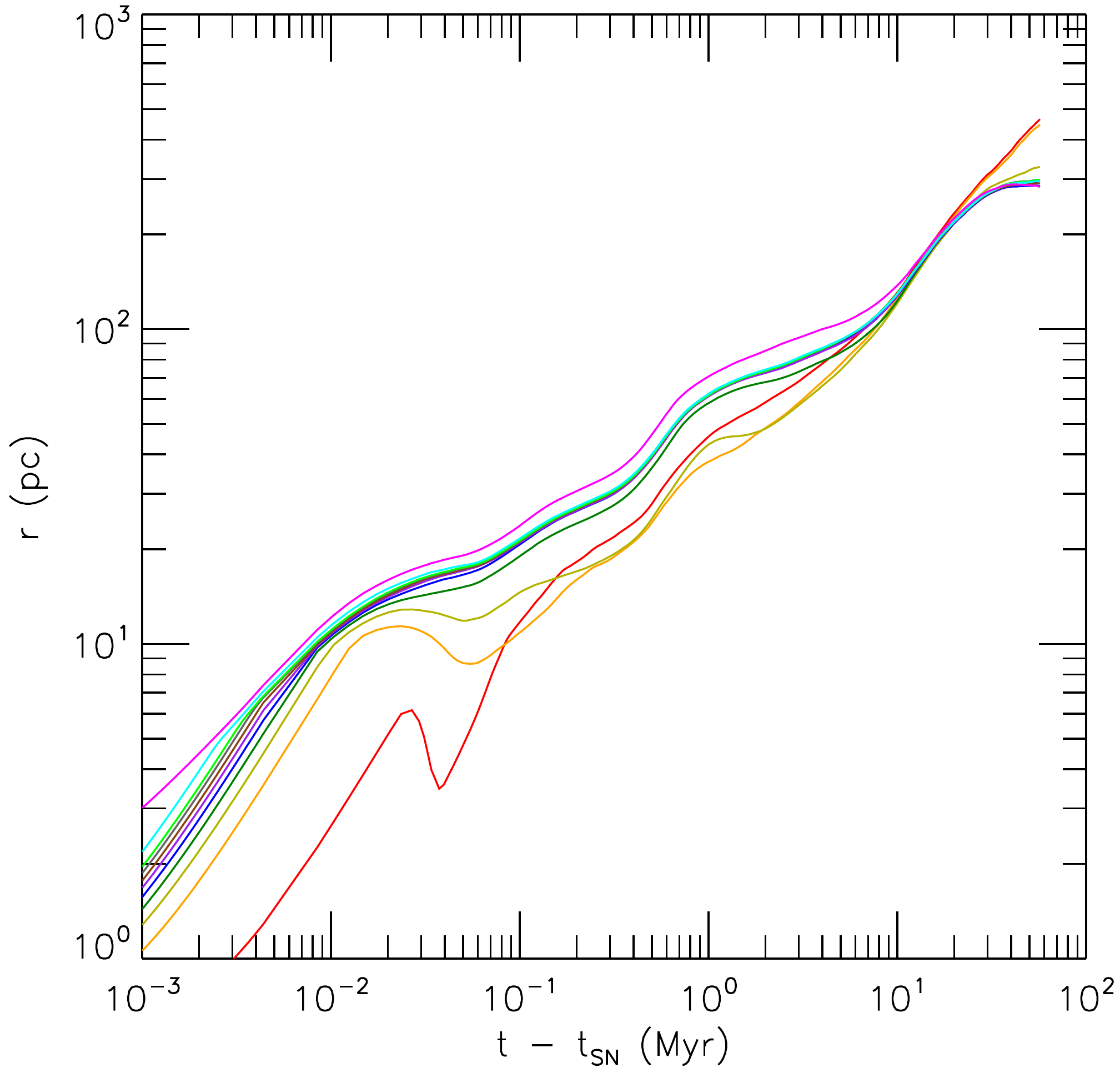}
\end{center}
\caption{The median radii of Lagrangian tracer particles originally at radial mass coordinates $<0.01\,M_{\rm ej}$ (red lines),  $0.1$, $0.2$, ..., $0.9\,M_{\rm ej}$ (orange to green line), and $>0.99\,M_{\rm ej}$ (purple line) as a function of time from SNR insertion.  Note the inversion in which inner shells overtake outer shells at late times.}
\label{fig:medianRadius}
\end{figure}

After the supernova remnant was inserted, the hydrodynamic time step was allowed to increase freely subject to the Courant-Friedrichs-Lewy criterion. The extent of the region forced to maximum refinement was progressively enlarged such that the entire SNR as identified by the extent of the forward shock remained at maximum refinement. The maximum refinement level $\ell_{\rm max}$ was allowed to decrease subject to the requirement that the diameter of the SNR be resolved by at least 128 cells. This meant that the maximum refinement changed to $\ell_{\rm max} = (12,\,10,\,9,\,8)$ at $t - t_{\rm SN} = (0.0044,\,0.18,\,1.2,\,8.7)\u{Myr}$, respectively, with $\ell_{\rm max}=8$ being the maximum refinement level at the end of the simulation.  At late times $t- t_{\rm SN}>8.7\u{Myr}$ when we focused on the gravitational fallback of metal-polluted gas to the center of the halo, we did not keep the entire remnant at the same refinement level, but instead refined the grid using the standard \textsc{flash} second-derivative-based refinement criterion applied to temperature, density, and metallicity.  The Jeans length in the metal-enriched gas was resolved by at least 16 cells until the end of the simulation.


\section{Results}
\label{sec:results}

\begin{figure*}
\begin{center}
\includegraphics[trim=0cm 0cm 0cm 6cm,clip=true,width=0.49\textwidth]{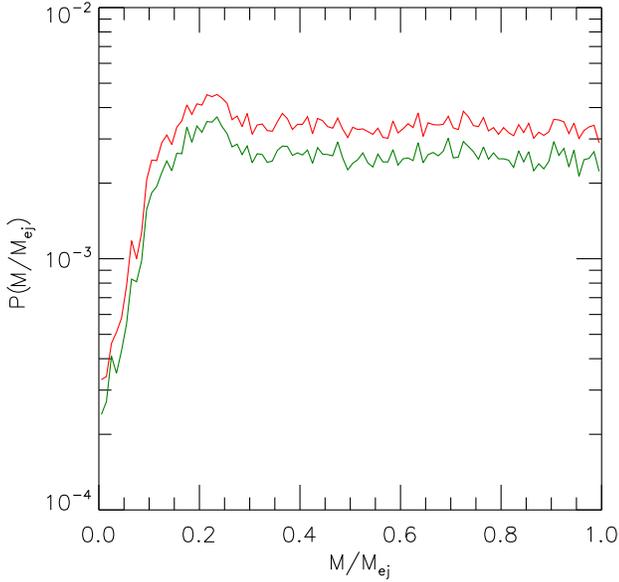}
\includegraphics[trim=0cm 0cm 0cm 6cm,clip=true,width=0.49\textwidth]{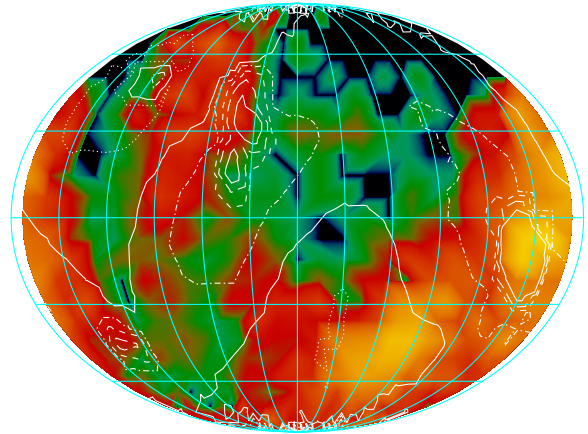}
\end{center}
\caption{Left panel: The probability that a particle with initial mass coordinate $M/M_{\rm ej}$ in the Pop III ejecta becomes incorporated in the central gas clump with density $\geq 10^{-22}\u{g}\u{cm}^{-3}$ (red line) and $\geq 10^{-21}\u{g}\u{cm}^{-3}$ (green line) that  is assumed to be capable of forming Pop II stars at the end of the simulation. Right panel: Mollweide projection of the probability of incorporation into the central gas clump with the density threshold of $10^{-21}\u{g}\u{cm}^{-3}$ as a function of the initial direction of ejection.  The probability increases logarithmically from $10^{-6}$ to $10^{-1}$ as the color ranges from black to yellow. The radially-averaged gas density in the annulus $10\u{pc}\leq r\leq 200\u{pc}$ at SNR insertion is shown in white contours; the contours are logarithmically spaced in density between $10^{-24.8}\u{g}\u{cm}^{-3}$ (dotted line) and $10^{-23.8}\u{g}\u{cm}^{-3}$ (solid line).}
\label{fig:probability}
\end{figure*}

\subsection{Morphological evolution of the supernova remnant}

Figure~\ref{fig:slice_projection} shows slices and projections of density and passive scalar metallicity at three different times after SNR insertion. In the left column showing the state of the remnant at $\approx 0.9\u{Myr}$ after the explosion, we see a thin snowplow shell enclosing a low-density  ($\sim 10^{-3}-10^{-2}\u{cm}^{-3}$) interior.  The elliptical features exterior to the snowplow shell are the shock wave driven outward by H\,II region pressure. The top-left-to-bottom-right diagonal feature is a gaseous filament of the cosmic web. Metals remain confined to the SNR shell and its interior.  

At $35\u{Myr}$ after explosion, hydrodynamic instability has already rendered the SNR highly anisotropic.  It has expanded to the box edges ($>500\u{pc}$) and has collapsed in the direction of the cosmic web filament. The supernova ejecta and entrained gas have begun to stream back into the halo center.  The densest gas is confined by the dark matter halo's gravitational pull and is not self-gravitating.  The peak gas density remains low, $\sim 1-10\u{cm}^{-3}$, insufficient to provide conditions for star formation.  Metal-carrying ejecta are confined in a thin, crumpled sheet perpendicular to the cosmic web filament. The  ejecta remain highly inhomogenous and undiluted: this may be less clear in the numerical-diffusion-susceptible passive scalar metallicity slice shown in Figure~\ref{fig:slice_projection}, but is visible in the Lagrangian tracer particles (not shown) that are distributed in thin sheets.

At $57\u{Myr}$ after explosion, the dark-matter-confined, turbulent, quasi-hydrostatic gas has reached density  $10^3\u{cm}^{-3}$.  The gas Jeans length has dropped to $\lambda_{\rm J}\sim 16\u{pc}\sim 16\,\Delta x$ and is decreasing rapidly.\footnote{The Jeans length is artificially large thanks to our neglect of H$_2$ cooling.}   Only a very small fraction of the ejecta has made it into the dense gas located at $(x,y)=(0,0)$ in the metallicity slice in Figure \ref{fig:slice_projection}. Dilution with metal-free gas has brought the metallicity in the densest gas down to below $10^{-4}\,Z_\odot$.

\begin{figure*}
\begin{center}
\includegraphics[width=\textwidth]{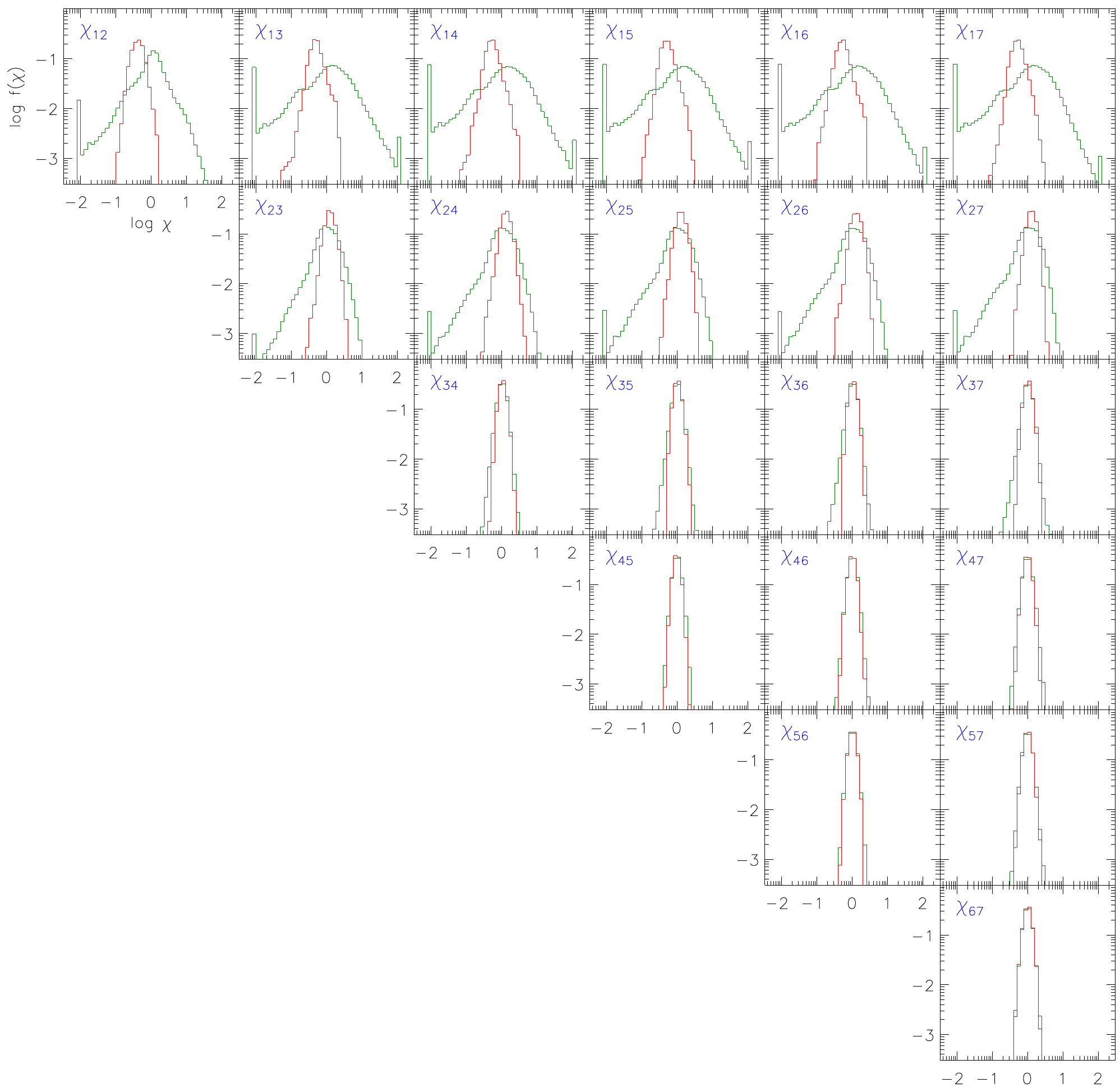}
\end{center}
\caption{Distributions of ejecta abundance fractions
  $\chi_{ij}=\rho_i/\rho_j$ at the end of the simulation, where $i,j=1,...,7$ label
  the radial mass bins at the time of SNR insertion. The histograms were computed with all Lagrangian
  tracer particles (green lines) and only particles within
  $50\,\textrm{pc}$ from the point of collapse to high densities (red
  lines). The local ejecta densities were smoothed employing an Epanechnikov kernel
   with kernel radius adjusted such that the combined,
  kernel-weighted metal mass, as defined in the text, was $2\times 10^{-4}\,M_\odot$.  Ejecta abundance ratios with $|\log\chi_{ij}|>2$
  were compressed into the flanking bins.}
\label{fig:histograms}
\end{figure*}

\subsection{Ejecta dispersal trends}
\label{sec:trends}

Figure~\ref{fig:medianRadius} shows the evolution of median radii of Lagrangian tracer particles as functions of the original radial mass coordinate in the freely-expanding SNR at insertion.  Apart from an early retrogression seen in the innermost shells $\sim 20-50\u{kyr}$ after SNR insertion that reflects reverse shock passage, all median Lagrangian radii are in expansion until $\sim 3\u{Myr}$.  Then the median radii of the envelope at mass coordinates $\gtrsim 0.3\,M_{\rm ej}$ begin to stall.  Interestingly, the median radii of the innermost mass coordinates corresponding to the core of the explosion continue to expand until the end of the simulation.  This inversion in the median Lagrangian radii in the inner 10\% of the ejecta already begins at $\sim1\u{Myr}$.  The inversion is a hydrodynamic consequence of post-reverse-shock entropy variation \citep{Ritter:15}.  This is most easily seen in the Sedov-Taylor idealization in which pressure tends to a constant value toward the center of the remnant whereas density tends to zero steeply with the ratio of the radius to the blastwave radius $\rho\propto (r/R_{\rm ST})^{3/(\gamma-1)}= (r/R_{\rm ST})^{9/2}$ \citep[see, e.g.,][\S106; the power is $9/2$ for non-relativistic electrons with adiabatic index $\gamma\approx 5/3$ and even steeper for relativistic electrons and $\gamma\approx 4/3$]{Landau:87}. Together, these scalings imply that specific entropy rises toward the center as $s\propto \ln(P/\rho^\gamma)+\mathrm{const} \propto - \frac{15}{2}\ln (r/R_{\rm ST})+\mathrm{const}$.  

We can treat the Sedov-Taylor radial coordinate $r/R_{\rm ST}\propto m^{2/15}$ as a function of the fraction $m$ of the swept up mass. Then the specific entropy varies with the mass coordinate as $s\propto -\ln m+\mathrm{const}$.  Hydrodynamic instability perturbs the remnant in the radial direction.  If the perturbation is adiabatic and the swept up gas remains in approximate pressure equilibrium ($P=\mathrm{const}$), the cooling time is proportional to $t_{\rm cool}\propto P/(\Lambda\rho^2)\propto P^{-1/5} e^{6s/5}/\Lambda\propto (P^{-1/5} /\Lambda)\,m^{-6/5}$, where $\Lambda$ is the usual temperature-dependent cooling coefficient in $\mathrm{erg}\,\mathrm{s}^{-1}\,\mathrm{cm}^{3}$. Since $\Lambda$ is not a strong function of temperature at SNR interior temperatures $\gtrsim 10^5\u{K}$, cooling time increases toward small mass coordinates. The gas enriched with the supernova progenitor's envelope cools faster than that enriched with the core.  The more rapidly cooling outer mass coordinates form a shell; fragments of this shell lose momentum as they run into dense surrounding clouds (here, cosmic web streams that have not been photoionized). Shocked gas elements that do not cool become upward buoyant when the remnant comes into pressure equilibrium with the ambient medium.  This, we believe, is the origin of the inversion in which the innermost mass coordinates of the ejecta travel farther than outer ones.

\begin{figure*}
\begin{center}
\includegraphics[trim=0cm 0cm 0cm 6cm,clip=true,width=0.35\textwidth]{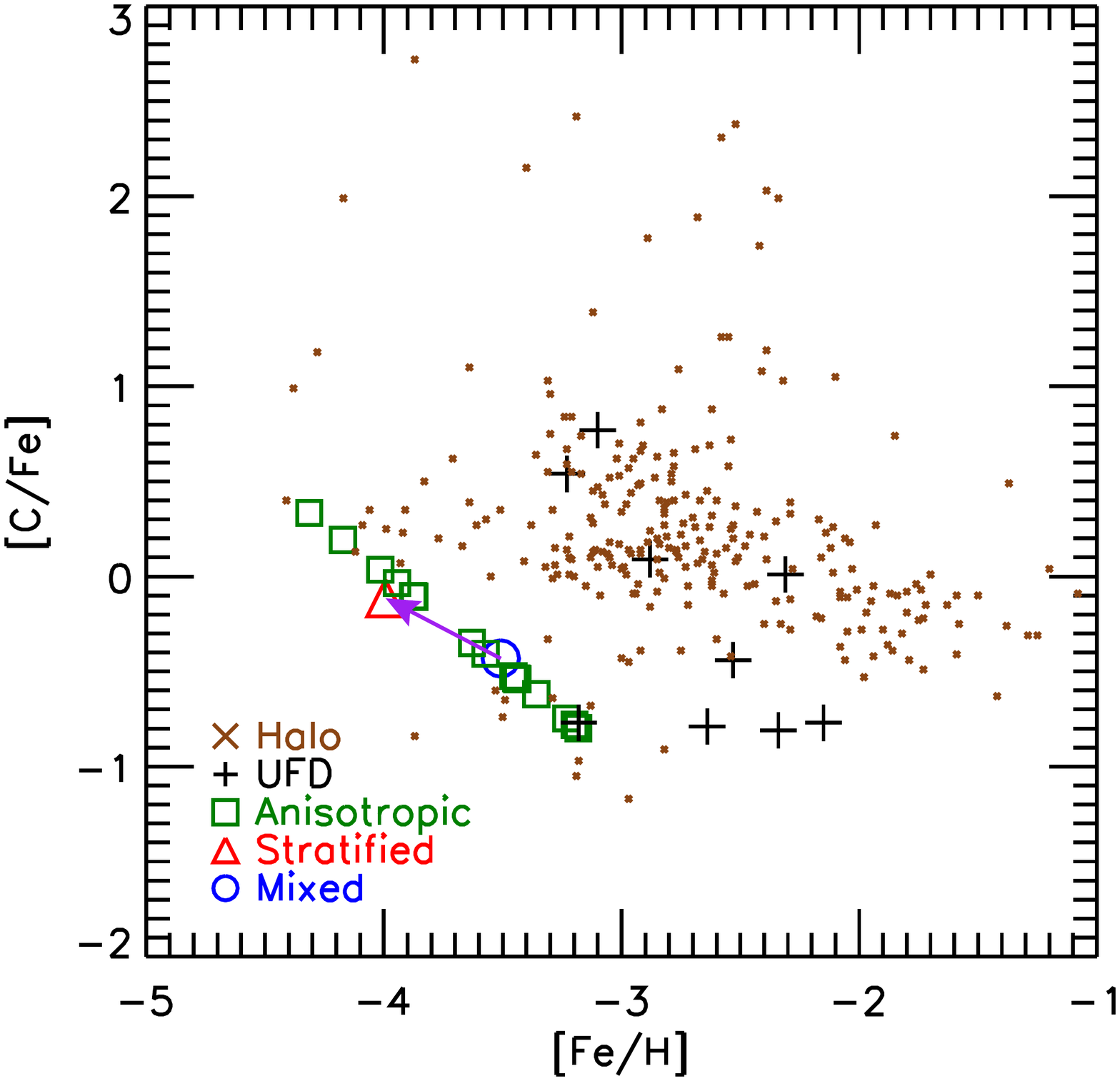}
\includegraphics[trim=0cm 0cm 0cm 6cm,clip=true,width=0.35\textwidth]{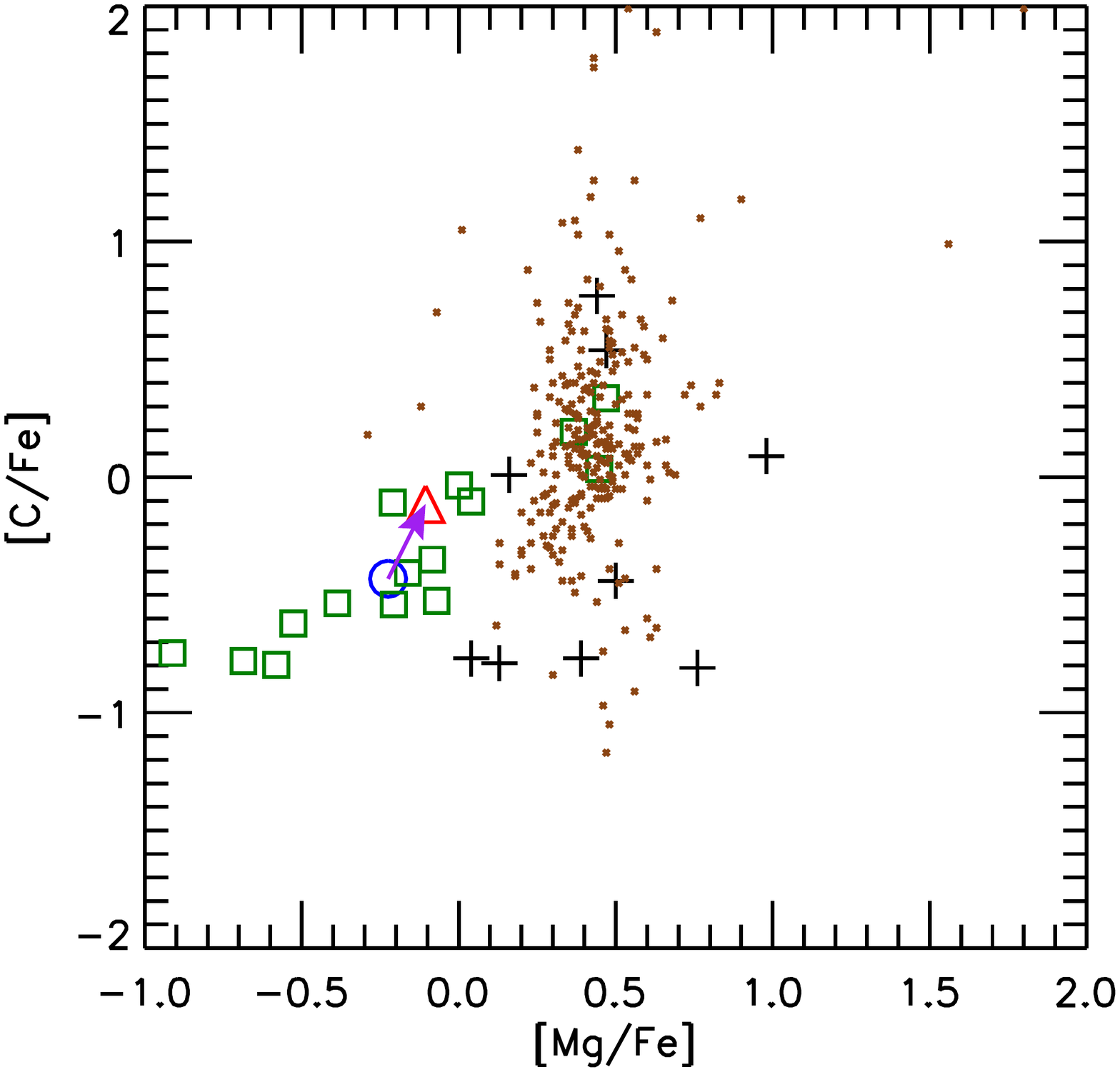}
\includegraphics[trim=0cm 0cm 0cm 6cm,clip=true,width=0.35\textwidth]{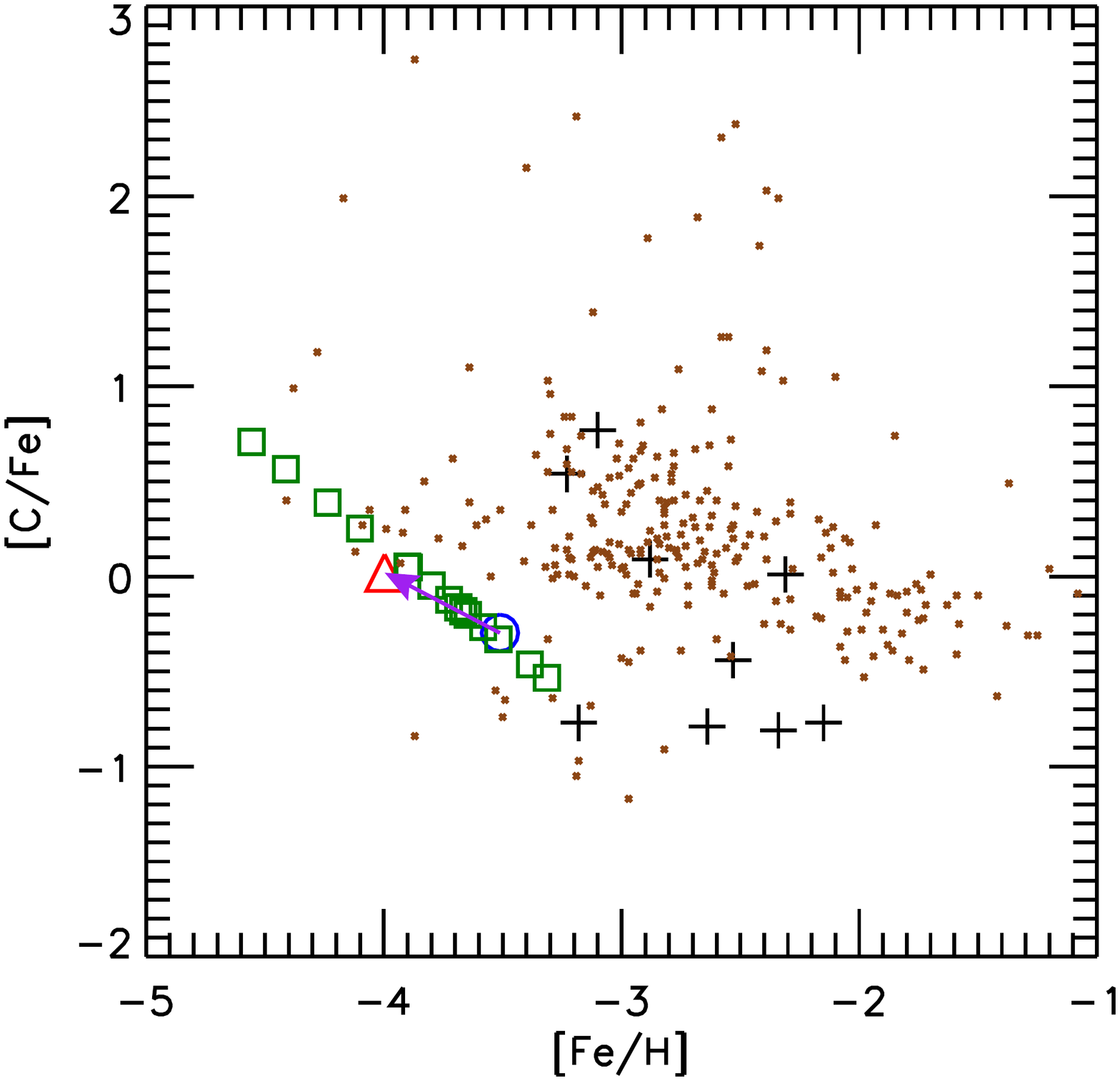}
\includegraphics[trim=0cm 0cm 0cm 6cm,clip=true,width=0.35\textwidth]{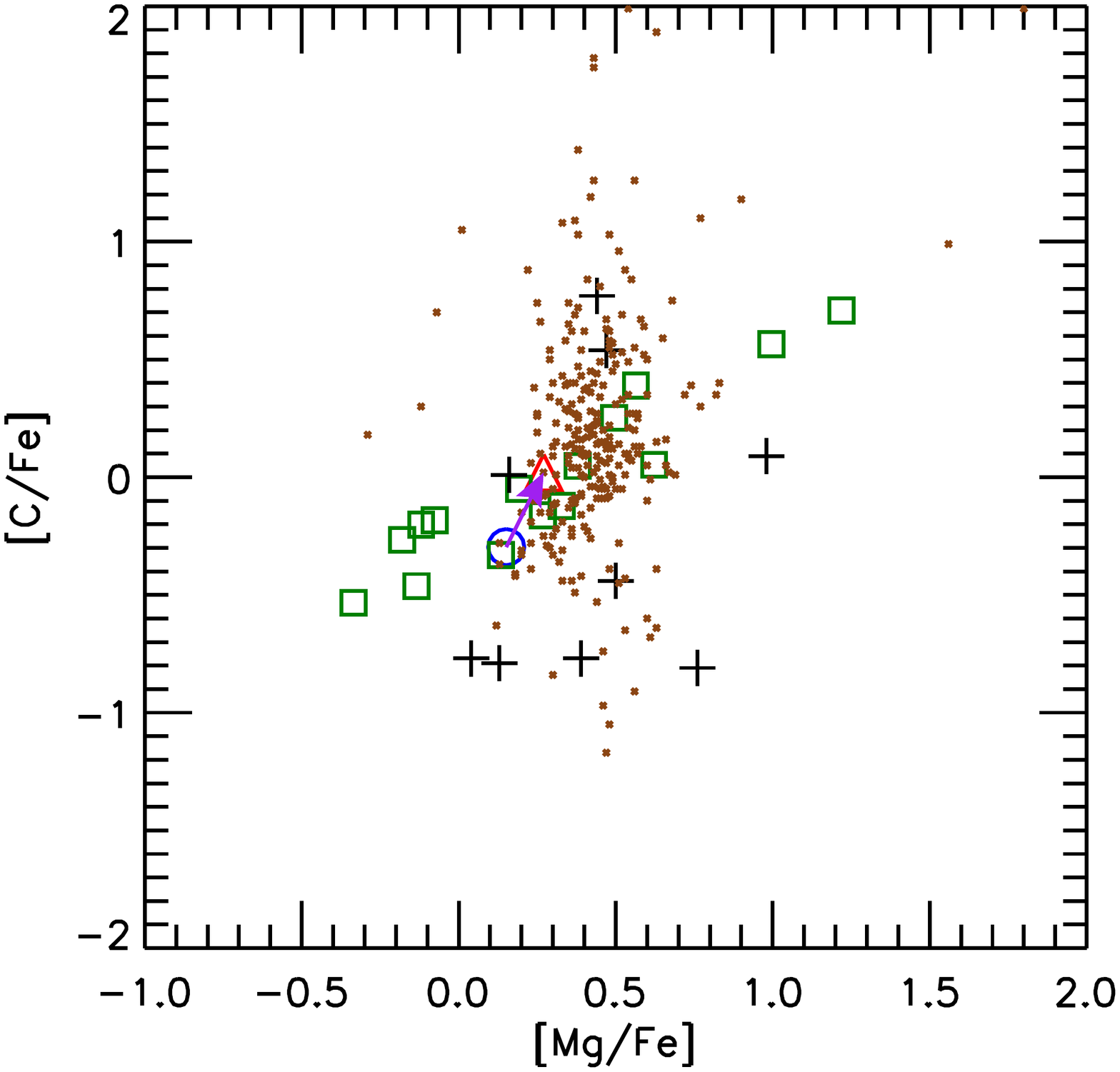}
\includegraphics[trim=0cm 0cm 0cm 6cm,clip=true,width=0.35\textwidth]{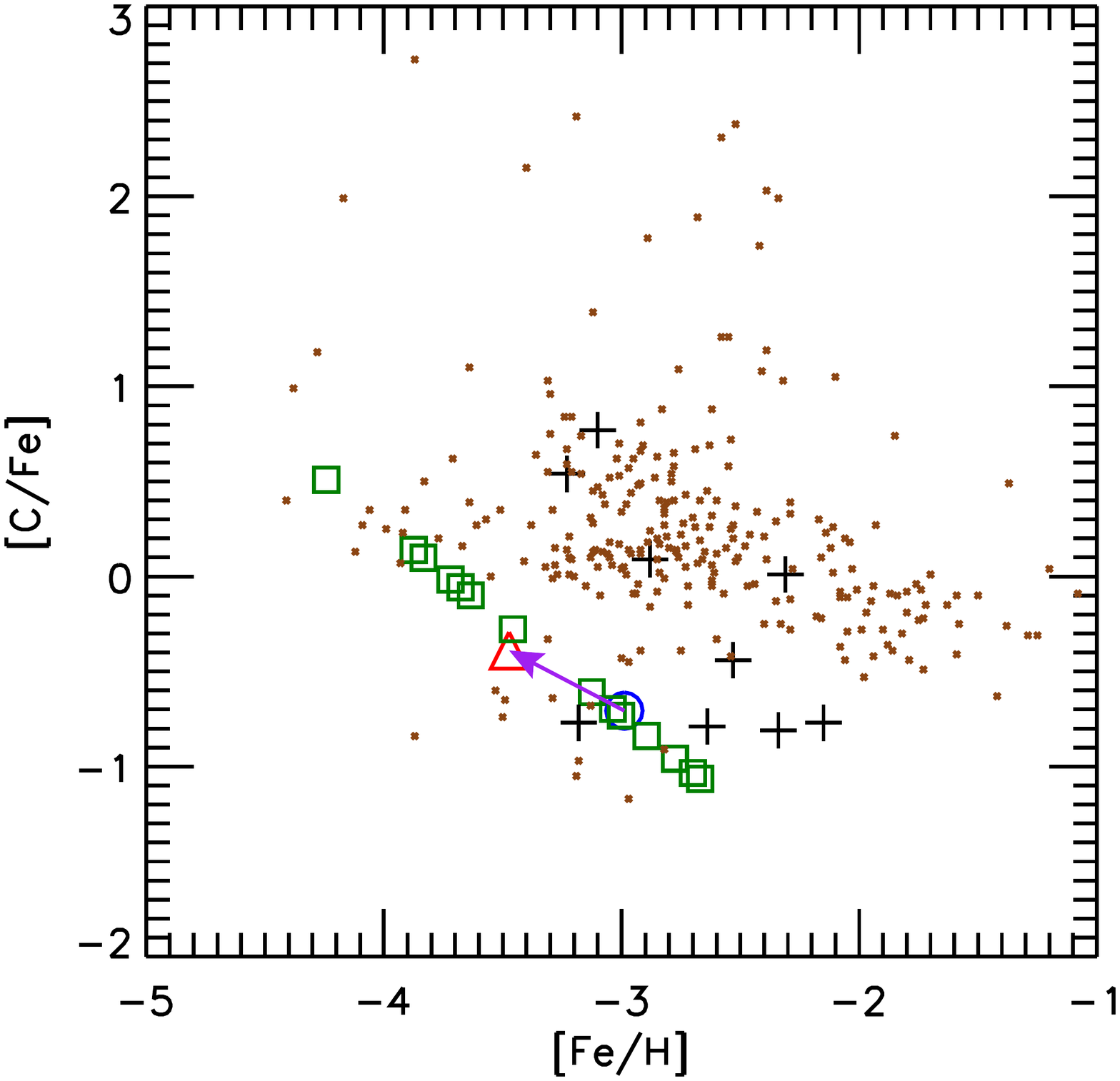}
\includegraphics[trim=0cm 0cm 0cm 6cm,clip=true,width=0.35\textwidth]{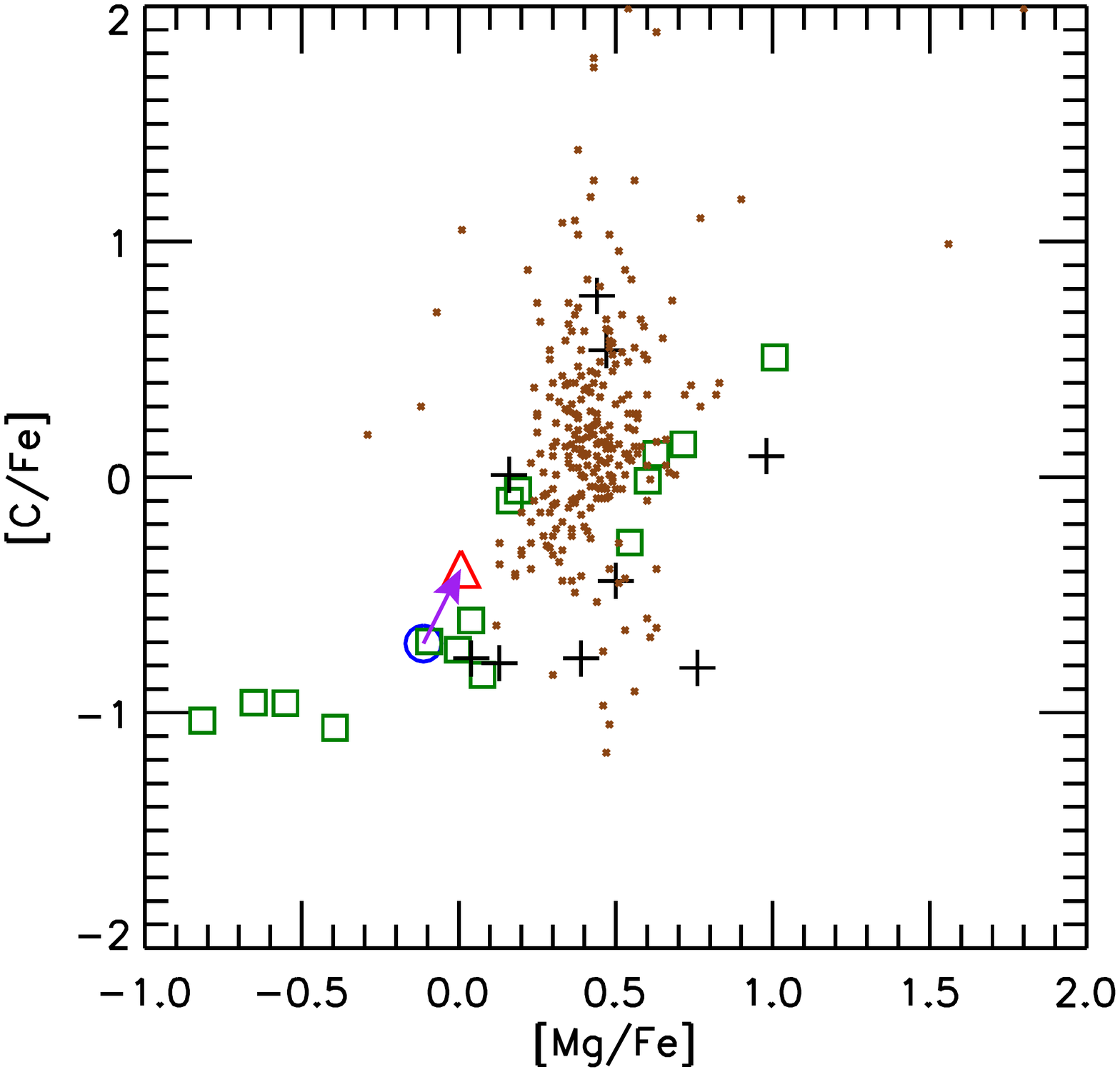}
\end{center}
\caption{Illustration of possible chemical space effects in Pop II stars resulting from chemical stratification or anisotropy in the Pop III supernova explosion.  The abundance ratios [Fe/H], [C/Fe], and [Mg/Fe] are compared for three types of artificial model explosions: completely mixed explosions (blue circles), spherically symmetric, radially stratified explosions (red triangles), and anisotropic explosions (green squares).  The yields of model explosions are calibrated, from top to bottom row, to $20$, $30$, and $40\,M_\odot$ models of \citet{Heger:10} with explosion energies $10^{52}\u{erg}$, mixing parameter $2.51\times10^{-3}$, and explosion mechanism S4 (ultra-energetic explosions were selected because spherically symmetric $10^{51}\,\mathrm{erg}$ explosions of Heger \& Woosley did not produce iron---iron ejection requires either broken spherical symmetry or very high temperatures).  For anisotropic explosions we emulated Rayleigh-Taylor fingers seen in the \citet{Wongwathanarat:15} core-collapse simulation: each element was shot into a discrete, random set of directions, 4 different directions for Fe and 12 directions for Mg.  Supernova models were mapped onto tracer particles at insertion and the particles in gas with density $\geq 10^{-21}\u{g}\u{cm}^{-3}$ at the end of the simulation were assumed to be the material from which a Pop II star cluster would form. The purple arrow shows the chemical space displacment from fully mixed ejecta to spherically symmetric, radially stratified ejecta.  The black $+$ symbols show observed abundance ratios in UFDs \citep{Frebel:10b,Ishigaki:14} and small brown $\times$ symbols show those in the Galactic stellar halo \citep{Roederer:14}.}
\label{fig:mainresult}
\end{figure*}

We turn to enrichment of the densest $\rho\geq 10^{-21}\u{g}\u{cm}^{-3}$ central gas clump that is poised to form second-generation stars. Figure~\ref{fig:probability}, left panel, shows the return probability, namely, the probability that a Lagrangian particle becomes incorporated in the dense clump, as a function of the radial mass coordinate in the ejecta.  The probability is markedly reduced for the inner $20\%$ of the ejecta.  This again is consistent with the hypothesis that the innermost ejecta are raised to high entropy, cannot cool, and because of upward buoyancy prefer to reside at larger radii.

Figure~\ref{fig:probability}, right panel, shows the return probability as a function of the initial direction in which the particle was traveling in the freely-expanding ejecta.  The return probability is very small in the darkest colored regions and higher by five orders of magnitude in the regions colored bright yellow (the color scale is logarithmic). Clearly the variation of the return probability in the ejecta direction is much more dramatic than in the ejecta radial mass coordinate. The white contours represent the mean density of gas in the annulus $10\u{pc}\leq r\leq 200\u{pc}$ as a function of direction. This is the density that the ejecta ``sees" as it expands into a highly anisotropic gaseous environment. The density maxima correspond to filaments of the cosmic web and minima correspond to voids. The return probability is correlated with the density in the sense that the Lagrangian tracer particles shot into directions with higher obstructing densities are more likely to become incorporated in the central gas clump.

\subsection{Chemical variation in diffuse and dense gas}
\label{sec:variation}

Having identified the gross chemical dispersal trends, we turn to the question, to what extent is the long-term gravitational hydrodynamics of a Pop III SNR able to erase an original chemical stratification in the supernova explosion?  In \citet{Ritter:15}, we found that the SNR remained chemically heterogeneous until a fraction of the enriched volume was gravitationally-compressed to densities $\gtrsim 10\u{cm}^{-3}$.  In this dense, gravitationally-collapsing clump, gravitational infall excited turbulence in which vorticity was high enough to ensure turbulent mixing and local chemical homogenization.  

Here, we carry the analysis further by deriving the full statistics of chemical abundance ratios. We divide the ejecta mass coordinate at insertion into 7 equal-mass bins.  At any point $\mathbf{r}$ we compute adaptive Epanechnikov kernel estimates of the local densities $\rho_i(\mathbf{r})$ and $\rho_j(\mathbf{r})$ of the material originating from bins $i$ and $j\leq i$ by choosing the kernel radius $R_{ij}({\mathbf r})$ such that the
combined kernel-weighted particle mass within the kernel footprint
equals a reference mass $M$,
\beq
\sum_{p\in {\mathcal Y}_i\cup {\mathcal Y}_j, |{\mathbf r}_p|<R_{\rm
    ij}({\mathbf r})} \left(1-\frac{|{\mathbf r}_p-{\mathbf
        r}|^2}{R_{\rm ij}({\mathbf r})^2}\right)  \,m_p = M .
\eeq
Here, ${\mathcal Y}_i$ is the set of Lagrangian tracer particles
originating from bin $i$ and ${\mathbf r}_p$ and $m_p$ are, respectively, the
position and mass of
particle $p$.  For the reference mass we choose
$M=2\times10^{-4}\,M_\odot=50\,M_{\rm part}$; this means that on average, $125$ particles are located in the kernel footprint.  Given this adaptive estimate of the kernel radius $R_{ij}({\mathbf r})$, we sample space with tracer particles in both bins, $i$ and $j$, and compute kernel-weighted estimates of the densities $\rho_i(\mathbf{r})$ and $\rho_j(\mathbf{r})$. We divide the two densities to obtain the ratios
\beq
\chi_{ij} = \frac{\rho_i}{\rho_j} .
\eeq
In Figure \ref{fig:histograms} we plot the probability density function (PDF) of $f_{ij}(\log\chi_{ij})$ for each pair of ejecta mass coordinate bins.  In addition, we compute  the PDF $f_{50,ij}(\log\chi_{ij})$ by restricting kernel estimate locations $\mathbf{r}$ to Lagrangian tracers that end up within $50\u{pc}$ from the halo center at the end of the simulation.

The spread in 
$f_{ij}$ decreases with increasing $i$, showing that the
outer, higher-initial-velocity mass shells of the ejecta are relatively well-stirred, with the small
spread consistent with Poisson sampling noise, but the
innermost mass shells, however, are poorly stirred with each other and the outer shells.  The distribution
exhibits a heavy tail deficient in inner shell ejecta (corresponding to low values of $\chi_{ij}$ for $i=1$ or $2$). 
 The halo center distribution $f_{50,ij}$ is more narrowly peaked at low
$i$ (inner shells) and exhibits a similar trend or decreasing spread with increasing ejecta mass coordinate. The displacement of the PDF peak toward low $\chi_{1j}$ shows that, consistent with the inner-ejecta-deficiency trend identified in Section \ref{sec:trends}, the central
$50\,\textrm{pc}$ is indeed deficient in the ejecta from the innermost mass coordinates of the explosion.

\subsection{Abundance biases and anomalies in metal-poor stars}

The question that remains is, how should the hydrodynamic trends and variation that we have identified in Sections \ref{sec:trends} and \ref{sec:variation} manifest in stellar chemical abundance spaces?  What is the pure effect of explosion geometry---when the yields are themselves held fixed---on the enrichment of second-generation stars?  Obviously, the single explosion that we have simulated cannot answer this general question.  We can still use the simulation as an indicator of what is the plausible magnitude of such effects.   For this purpose, we construct artificial model explosions: a fully mixed explosion, a spherically-symmetric, radially-stratified explosion, and an anisotropic explosion.  In search of a suite of template Pop III star explosion models, we sought such models in the database of one-dimensional piston-triggered explosions of \citet{Heger:10}.  We needed models that produced at least some iron, but their canonical-energy ($10^{51}\u{erg}$) explosions did not as all iron group elements were lost in the compact remnant.  This is of course an artifact of spherical symmetry.  

To understand how spherical symmetry breaks down, consider the iron core collapse.  The flow onto the proto-neutron star rebounds and breaks into an accretion shock. Energy loss to nuclear dissociation and neutrino emission stalls the shock, but neutrinos from the neutron star heat the shocked, infalling matter and produce a strong negative radial entropy gradient. The gradient implies vigorous turbulent convection.  The largest convective plumes reach the standing, already aspherical accretion shock, and in some directions push the shock outward, into the envelope of the star.  The matter in the plumes is dissociated into $\alpha$-particles and they recombine into $^{56}$Ni as the plumes rise and cool. Nickel is also synthesized in explosive burning of pre-supernova silicon \citep[see, e.g.,][]{Kifonidis:03}.  Buoyancy accelerates the under-dense, nickel-rich plumes to high velocities. When the plumes cross the interface between the C-O and He shells, a hydrodynamic process akin to the Rayleigh-Taylor instability stretches them into long, narrow fingers. Some three-dimensional neutrino-driven, core-collapse explosions of \citet{Wongwathanarat:13,Wongwathanarat:15} and \citet{Utrobin:14} that track the explosions until shock breakout develop prominent fingers with amplitudes increasing with decreasing mass coordinate.  The most prominent $^{56}$Ni fingers can outrun the weaker $^{16}$O, $^{20}$Ne, and $^{24}$Mg fingers, and both can transport the heavier elements ahead of the isotropically-ejected shell of $^{12}$C. This ensures iron group escape.  

It should be emphasized that the production of nickel fingers is sensitive to the precise shock kinematics inside the star; not all stellar models in \citet{Wongwathanarat:15} produce fingers. When the forward shock crosses the He-H interface (if the progenitor star has retained its hydrogen envelope), it slows down dramatically, and this excites a reverse shock propagating backward, toward smaller mass coordinates. The reverse shock in turn decelerates any outward-traveling material it crosses. If a finger is slower than and is trailing the forward shock, as in the $20\,M_\odot$ blue supergiant model in Wongwathanarat et al., then it collides with the reverse shock.  This compresses and potentially even disrupts the finger. On the other hand, if the finger is faster than and is leading the forward shock, as in the $15\,M_\odot$ blue supergiant model, nickel is already ahead when the reverse shock forms and is not affected.  The nickel travels ballistically toward the stellar surface and beyond. The finger evolution becomes more complicated in less clear-cut cases, such as in both of Wongwathanarat et al.'s red supergiant models.

In view of this multidimensional complexity and the limitations it implies for the robustness of the Heger \& Woosley models, for our analysis we artificially adopt their ultra-energetic ($10^{52}\u{erg}$) explosion yield patterns that \emph{did} produce iron, albeit in small quantities.  Specifically, we select three models with progenitor masses of $20$, $30$, and $40\,M_\odot$ and radial stratification as in Figure 7b of Heger \& Woosley.  For anisotropic explosions, we mimic, in a statistical fashion, the fingering as seen in the \citet{Wongwathanarat:15} simulations.  Emulating their Figure 3, we take there to be 4 Fe fingers and 12 Mg fingers.  The radial extent of the carbon shell was $0.1-0.15\,M_{\rm ej}$ in the ejecta mass coordinate. The angular diameters of the Fe and Mg fingers were $20^\circ$ and $14^\circ$, respectively, and their mass coordinate extents were $0-0.3\,M_{\rm ej}$ and $0.05-0.2\,M_{\rm ej}$.

We use the template models to assign chemical identity to Lagrangian tracer particles based on their location in the SNR at insertion.  Then we study chemical abundance ratios [Fe/H], [C/Fe], and  [Mg/Fe] in the tracer particles that have returned to the halo center to become incorporated in the dense $\geq 10^{-21}\u{g}\u{cm}^{-3}$ clump.  In anisotropic explosions, we generate several realizations by shooting fingers in random directions.  The resulting abundance ratios are shown in Figure~\ref{fig:mainresult}.  
All models predict values of [Fe/H] that are too low compared to observations, but this is again an artifact of the loss of iron to the compact remnant in the spherically-symmetric computations of Heger \& Woosley.  The chemical space displacement from completely mixed to stratified models is 
\begin{eqnarray}
\Delta [\mathrm{Fe}/\mathrm{H}] &\approx& -0.5 ,\nonumber\\
\Delta [\mathrm{C}/\mathrm{Fe}] &\approx& +0.3 ,\nonumber\\
\Delta [\mathrm{Mg}/\mathrm{Fe}] &\approx& +0.1 ,
\end{eqnarray}
and the displacements in anisotropic models lie in the interval
\begin{eqnarray}
(-1.2,\,-0.8) &\lesssim& \Delta [\mathrm{Fe}/\mathrm{H}] \ \ \ \lesssim \ \ \ (+0.2,\,+0.3) ,\nonumber\\
(-0.4,\,-0.2)&\lesssim& \Delta [\mathrm{C}/\mathrm{Fe}] \ \ \ \lesssim \ \ \ (+0.7,\,+1.2) ,\nonumber\\
(-0.7,\,-0.5) &\lesssim& \Delta [\mathrm{Mg}/\mathrm{Fe}]\ \ \  \lesssim \ \ \ ( +0.7,\, +1.1),
\end{eqnarray}
where the values in parentheses reflect model-to-model variation.  The variation in anisotropic models comes close to reproducing the scatter in the observed UFDs and the Galactic halo.

\section{Discussion and conclusions}
\label{sec:discussion_conclusions}

Previous studies of chemical abundance pattern formation in the first Pop II stars \citep[e.g.,][]{Karlsson:13,Ishigaki:14,Cooke:14,Kobayashi:14,Tominaga:14} have assumed that individual Pop III supernovae contribute their nucleosynthetic yields ``monolithically'': the abundances observed in the most metal-poor stars are linear combinations of the integral yields of the contributing supernovae. This approach rests on the assumption that the supernova ejecta is completely mixed either in the explosion, or in the SNR, or perhaps in the larger confining gravitational potential well of the host dark matter halo. Here and in \citet{Ritter:15}, we have provided counterexamples to this assumption, all deriving from the inevitable observation that the spatial distribution and kinematic state of the heavier elements returned by the explosion are highly inhomogeneous and anisotropic and that the long-term SNR hydrodynamics can partially preserve such chemical heterogeneity.\footnote{We have focused on neutrino-powered explosions.  It is worth separately examining consequences of explosion anisotropy in the distinct, hypothetical class of supernovae powered by the rotational energy of the central compact object and a magnetized inflow/outflow process.}  

One mechanism that we have identified is that the inner ejecta cool less efficiently than the outer ejecta due to the inverse relationship between the radius and the amount of specific entropy creation in the passage of the reverse shock. This mechanism acts to preserve some of the initial radial stratification of the ejecta in the material returning to the halo center to form second-generation stars. The other mechanism relates to angular anisotropy of radial deceleration as ejecta collide with an anisotropic distribution of dense clouds surrounding the explosion.  This mechanism preserves angular variation in chemical composition of the ejecta, the variation that could arise from, e.g., Rayleigh-Taylor fingering preceding supernova shock breakout.  Both mechanisms act to bias stellar enrichment against the gross yields of the contributing supernovae.  They can also in themselves contribute to the scatter in abundance ratios, potentially explaining a fraction of the observed scatter in the most metal-poor stars (most of the scatter is a consequence of the stochastic nature of the formation of and nucleosynthesis in Pop III and early Pop II stars). Thus, in the initial stage of chemical enrichment, we predict an irreducible degree of abundance scatter, either from  anisotropy in a single explosion (this study), or from entropy effects in a small multiple of contributing supernovae \citep{Ritter:15}.

In conclusion, inspired by recent three-dimensional simulations showing extreme inhomogeneity and anisotropy in neutrino-powered core-collapse supernovae \citep{Wongwathanarat:13,Wongwathanarat:15,Utrobin:14}, as well as by the growing evidence that the most metal-poor stars exhibit abundance anomalies \citep{Feltzing:09,Cohen:13,Yong:13} not expected for full mixing of supernova yields, we carried out a simulation of the long-term transport of nucleosynthetic products from a Pop III supernova explosion until ejecta-enriched gas has recollapsed into into a dense, Pop II-star-forming clump.  Tracing elements back to their specific locations in the explosion and mapping onto inhomogeneous explosion models, we modeled the scatter in the abundance ratios of the most metal-poor stars. A proof of principle rather than a predictive model, our computation shows that SNR hydrodynamic effects can contribute to the large scatter in [C/Fe] and [Mg/Fe] in the most metal-poor stars.  In view of the accelerating stellar chemical abundance data gathering with large spectroscopic surveys such as HERMES-GALAH \citep{Zucker:12,DeSilva:15} and Gaia-ESO \citep{Gilmore:12} as well as the recent progress in {\it ab initio} supernova modeling \citep[for a review, see][and references therein]{Mezzacappa:15}, Lagrangian tagging of the small scale structure of supernova ejecta in simulations of cosmic chemical enrichment and star formation seems like a particularly promising tool for future inquiry.

\section*{Acknowledgments}

We acknowledge helpful conversations with Sean Couch, Anna Frebel, John Scalo, and Craig Wheeler.
The \textsc{flash} code was in part developed by the DOE-supported Flash Center
for Computational Science at the University of Chicago. The authors
acknowledge the Texas Advanced Computing Center at The University of
Texas at Austin for providing HPC resources under XSEDE allocation
TG-AST120024. CSS is grateful for support provided by the NASA Earth
and Space Science Fellowship (NESSF) program. This study was supported
by the NSF grants AST-1009928 and AST-1413501 and by the NASA grant NNX09AJ33G. Some of of the
visualizations were made with the \textsc{yt} package \citep{Turk:11}.

\footnotesize{

}

\label{lastpage}

\end{document}